\documentclass[11pt,a4paper]{article}
\usepackage{graphicx}
\usepackage{epstopdf}

\newcommand{\beq}{\begin{equation}}
\newcommand{\eeq}{\end{equation}}
\newcommand{\beqa}{\begin{eqnarray}}
\newcommand{\eeqa}{\end{eqnarray}}

\def\apd{{}_{(d)}}
\def\ape{{}_{(e)}}

\def\rgdb{\sqrt{ \apd \bar g }}

\def\rge{\sqrt{ \ape g }}

\def\D{\mathcal{D}}

\def\rgd{\sqrt{ \apd g }}
\def\Box{\nabla^2}      
\def\rgb{\sqrt{ \bar g }}
\def\apd{ {}_d } 
 
\def\ape{ {}_e }

\def\bpde{\bar \partial}
\def\bnabla{\bar \nabla}

\def\bpde{\bar \partial}
\def\bnabla{\bar \nabla}

\usepackage{feynmf}

\usepackage{adjustbox}
\usepackage{standalone}
\usepackage{amsmath}

\usepackage{xcolor}
\usepackage{epsfig}
\usepackage{cancel}
\usepackage{caption}
\usepackage{feynmf} 
\usepackage{amssymb}
\usepackage{amsfonts}
\usepackage{epsf}
\usepackage{rotating}

\usepackage{amsmath}
\usepackage{fancyhdr}

\usepackage{pstricks}
\usepackage{color}
\usepackage{frontespizio}
\usepackage{hyperref}
\hypersetup{
	colorlinks,
	citecolor=green,
	filecolor=black,
	linkcolor=blue,
	urlcolor=black
}
\usepackage{type1ec}
\usepackage[T1]{fontenc}
\usepackage{lettrine}
\usepackage{bbold}
\usepackage{calligra}
\usepackage{tikz}

\usepackage{mathrsfs}
\usepackage{curve2e}
\usepackage{setspace}
\usepackage{indentfirst}
\usepackage{emptypage}
\usepackage{cite}
\usepackage[font=small,labelfont=bf,labelsep=quad]{caption} 
\usepackage{graphicx} 
\usepackage{listings} 
\usetikzlibrary{patterns}
\usepackage{relsize}
\usetikzlibrary{intersections,positioning}
\usetikzlibrary{decorations.pathmorphing,decorations.markings,arrows,positioning}
\usepackage{braket}
\usepackage{mathrsfs}
\usepackage{calc}
\newlength\shlength
\newcommand\xshlongvec[2][0]{\setlength\shlength{#1pt}%
	\stackengine{-5.6pt}{$#2$}{\smash{$\kern\shlength%
			\stackengine{7.55pt}{$\mathchar"017E$}%
			{\rule{\widthof{$#2$}}{.57pt}\kern.4pt}{O}{r}{F}{F}{L}\kern-\shlength$}}%
	{O}{c}{F}{T}{S}}

\newcommand{\cS}{\mathcal{S}}

\newcommand{\rg}{\sqrt{g}}
\newcommand{\f}{\frac{1}{2}}
\newcommand{\de}{\delta}

\newcommand{\dfun}[2]{ \frac{\delta #1}{\delta #2}}
\def\lq{\left[}
\def\rq{\right]}
\def\lt{\left(}
\def\rt{\right)}

\newcommand{\sdfrac}[2]{\mbox{\small$\displaystyle\frac{#1}{#2}$}}

\IfFileExists{dsfont.sty}
{\usepackage{dsfont}
	\let\mathbb=\mathds
	\newcommand{\id}{\mathds{1}}}
{\typeout{Package dsfont.sty was not found, using alternative macros.}
	\let\mathds=\mathbb
	\newcommand{\id}{\mbox{1 \kern-.59em {\rm l}}}}

\usepackage{slashed}
\usepackage{units}
\usepackage{setspace}
\newcommand{\nn}{\nonumber}
         \let\d=\delta
         
        \let\m=\mu
\let\n=\nu      \let\x=\xi           
        
     \let\y=\psi    
\let\D=\Delta
\let\d=\delta

\newcommand{\sm}{\mathcal{S}}
\newcommand{\secref}[1]{Section~\ref{#1}}		
\newcommand{\pa}{\partial}						
\newcommand{\del}{\delta}

\newcommand{\vf}{\phi}

\def\nbox#1#2{\vcenter{\hrule \hbox{\vrule height#2in
			\kern#1in \vrule} \hrule}}
\def\sq{\,\raise.5pt\hbox{$\nbox{.09}{.09}$}\,}
\def\sqb{\,\raise.5pt\hbox{$\overline{\nbox{.09}{.09}}$}\,}
\newcommand{\tphi}{\tilde{\phi}}
\newcommand{\na}{\nabla}

\newcommand{\bea}{\begin{eqnarray}}
\newcommand{\eea}{\end{eqnarray}}
\newcommand{\be}{\begin{equation}}
\newcommand{\ee}{\end{equation}}

\newcommand{\bes}{\begin{subequations}}
	\newcommand{\ees}{\end{subequations}}

\def\nn{\nonumber\\}

\def\Box{\sq}

\numberwithin{equation}{section}

\usepackage{accents}

\makeatletter
\newcommand{\xLine}[2][]{\ext@arrow 0359\Rightarrowfill@{#1}{#2}}
\makeatother
\xdefinecolor{darkgreen}{RGB}{102, 204, 70}
\xdefinecolor{darkblue}{RGB}{0, 0, 153}
\usepackage{amssymb}
\usepackage{pifont}
\begin{document}

\begin{center}
\vspace{1.5cm}
{\Large \bf  Topological Corrections and Conformal Backreaction in the Einstein Gauss-Bonnet$/$Weyl Theories of Gravity at $D=4$
\\}
\vspace{1.5cm}
\vspace{0.1cm}
 \vspace{0.3cm}
\vspace{1cm}
{\bf $^{(1)}$Claudio Corian\`o, $^{(2,3)}$Matteo Maria Maglio and $^{(1)}$Dimosthenis Theofilopoulos\\}
\vspace{1cm}
{\it  $^{(1)}$Dipartimento di Matematica e Fisica, Universit\`{a} del Salento \\
and INFN Sezione di Lecce, Via Arnesano 73100 Lecce, Italy\\}
\vspace{0.5cm}
{\it  $^{(2)}$Galileo Galilei Institute for Theoretical Physics, \\
	Largo Enrico Fermi 2, I-50125 Firenze, Italy\\
	and\\
	$^{(3)}$Institute for Theoretical Physics (ITP), University of Heidelberg\\
	Philosophenweg 16, 69120 Heidelberg, Germany}

\begin{abstract}
We investigate the gravitational backreaction, generated by coupling a general conformal sector to external, classical gravity,  as described by a conformal anomaly effective action. We address the issues raised by the regularization of the topological Gauss-Bonnet and Weyl terms in these actions and the use of dimensional regularization (DR).    
We discuss both their local and nonlocal expressions, as possible IR and UV descriptions of conformal theories, below and above the conformal breaking scale. Our discussion overlaps with several recent studies of dilaton gravities - obtained via a certain singular limit of the Einstein-Gauss-Bonnet (EGB) theory - originally introduced as a way to bypass Lovelock's theorem. We show that nonlocal, purely gravitational realizations of such EGB theories, quadratic in the dilaton field, beside their local quartic forms, are possible, by a finite renormalization of the Euler density.
Such nonlocal versions, which are deprived of any scale, can be expanded, at least around flat space,   
in terms of the combination $R \Box^{-1}$ times multiple variations of the anomaly functional, as pointed out in recent studies at $d=4$. Similar conclusions can be drawn for the proposed nonlocal EGB theory. The expansion emerges from previous investigations of the anomalous conformal Ward identities that constrain such theories around the flat spacetime limit in momentum space.

\end{abstract}

\end{center}
\newpage

\section{Introduction} 
The search for corrections to general relativity (GR) and to its Einstein-Hilbert (EH) action by higher derivative terms, is characterized by a large number of both older and of more recent proposals. Their goal  is to address unsolved 
issues, such as the nature of dark energy \cite{Antoniadis:2011ib} and the mechanism of inflation of the early universe \cite{Starobinsky:1980te}, in a 
more satisfactory way.\\
 From $f(R)$ theories to models incorporating a dilaton field (dilaton gravities), including Horndeski and Lovelock actions \cite{Charmousis:2014mia}, just to mention a few, important issues  need to be addressed both of phenomenological and of theoretical character.  
An important open question concerns the quantum consistency of these extensions, since the presence of higher order derivatives in the action leads, in general, to equations of motion of higher order. \\
In quantum gravity, particular attention is paid to the stability and the consistency of such theories, by showing, for instance, the absence of tachyonic solutions as well as of ghosts and, eventually, addressing their renormalizability \cite{Hamada:2014pba} in a perturbative context. \\
Among these proposals, of particular interest are those extensions that lead to second-order equations of motion, even though they are generated by Lagrangians with higher derivatives. Such Lagrangians may be introduced at classical level, or, alternatively, they may originate from the inclusion of quantum corrections, in models where gravity is still treated classically. Their structure depends on the specific type of matter sector that is integrated out of the quantum partition function. If the matter sector is conformal, we will refer to the ensuing semiclassical effective action as an action modified by a conformal backreaction. \\
In theories of induced gravity, the partial integration - in the partition function - of a matter sector, can be sufficient, just by itself, to recover a EH action for gravity, accompanied by extra, higher derivative terms. Both the $R^2$ corrections and the EH term can be generated this way, realizing Sakharov's proposal of induced gravity \cite{Visser:2002ew}. In this case, the spacetime is a Lorentzian manifold and the metric is essentially free, while its dynamics is entirely induced by the inclusion of quantum corrections due to a generic matter sector.  These induce an effective action of the form
\beq
\sm=\int d^4 x \sqrt{g}\left( c_1 \frac{M_P^2}{2}R + c_2 R^2 +\ldots   \right),
\eeq
where the ellipsis refer to extra contributions built out of higher order geometrical invariants and $c_1, c_2$ are numerical constants.
In this approach, the entire gravitational theory can be viewed as the result of the quantum backreaction on a freely fluctuating metric, induced by the path integration over the matter sector. For a generic matter sector 
(non conformal) the effective action will include the dimensionfull constants of the theory, combined with the size of the extra dimensions, that acts as a second scale.\\
There are drastic simplifications if the matter sector is conformal \cite{Codello:2012sn,Coriano:2013xua,Coriano:2013nja}. 
The integration of a conformal sector allows to derive a form of gravity which is expressed uniquely in terms of corrections extracted from the two invariants $V_E$ and $V_{C^2}$, defined in terms of the Euler density and the square of the Weyl tensor.\\
 Around flat space, the corresponding effective action is characterized only by two scales, the renormalization scale $\mu$ introduced to regulate the UV behaviour of the theory, and the IR scale (here denoted by $L$) coming from the extra dimensions (ED). Such dependences, in this case, are only logarithmic.  
 
\subsection{Content of this work}
The goal of this work is to investigate the relation between anomaly actions, in their local and nonlocal formulations, and the $4d$ Einstein-Gauss Bonnet (EGB) theory, related to the inclusion of $V_E$, that has been extensively discussed in recent studies \cite{Glavan:2019inb,Anastasiou:2020zwc,Matsumoto:2022fln}. Theories of both types share some of their features, but also exhibit substantial differences, that we are going to highlight. Indeed, borrowing from previous results on conformal anomaly actions, we show that one can identify a nonlocal version of 
the $4d$ EGB theory. It differs from former formulations of such a theory, now recognized as dilaton gravities, for being nonlocal. The wide interest towards this new form of dilaton gravity \cite{Hennigar:2020lsl,Fernandes:2020nbq,Easson:2020mpq,Kobayashi:2020wqy,Konoplya:2020qqh,Bonifacio:2020vbk,Ai:2020peo,Wei:2020ght,Aoki:2020lig,Nojiri:2020tph,Konoplya:2020bxa,Guo:2020zmf,Fernandes:2020rpa,Casalino:2020kbt,Hegde:2020xlv,Ghosh:2020vpc,Doneva:2020ped,Zhang:2020qew,Konoplya:2020ibi,Singh:2020xju,Ghosh:2020syx,Konoplya:2020juj,Kumar:2020uyz,Zhang:2020qam,HosseiniMansoori:2020yfj,Wei:2020poh,Singh:2020nwo,Churilova:2020aca,Islam:2020xmy,Mishra:2020gce,Konoplya:2020cbv,Zhang:2020sjh,EslamPanah:2020hoj,Aragon:2020qdc,Aoki:2020iwm,Shu:2020cjw,Mahapatra:2020rds,Lu:2020iav,Gurses:2020ofy,Banerjee:2020dad,Ge:2020tid,Yang:2020jno,Lin:2020kqe,Yang:2020czk}, derived from a specific regularization of the topological GB term, motivates our comparative analysis.
\section{EGB theories with a  singular limit}
$4d$ EGB theories are generated by performing a singular limit on the coupling constant of the topological Gauss-Bonnet term, which is deprived of any dynamical content in $d=4$, but  not so after an infinite renormalization of the coupling. A finite action is generated by performing the $d\to 4$ limit of this term, exploiting its evanescence 
in the equations of motion of the metric. By a careful analysis, one derives a $(0/0)$ contribution to the classical action which includes both gravity and a dilaton field. It allows to define a theory of dilaton gravity which is quartic in the dilaton field. As we are going to show, this is not the only possibility. \\
 In our case, borrowing from previous analysis in the literature on anomaly actions and the inclusion of a finite renormalization of the topological GB term, 
we show that one can define a $4d$ EGB action which is quadratic in the dilaton field, and can be rewritten in a nonlocal form, by solving for the same field in terms of the metric. With no surprise, the action takes the form originally introduced by Riegert in the search of a functional solution of the anomaly constraint, the anomaly induced action, which was directly investigated at $d=4$ and not in a context of dimensional regularization (DR). \\
We remind that the analysis of anomaly induced actions take the form of  searches of solutions of anomaly constraints with no reference to the Weyl invariant terms coming from the virtual corrections.  For this reason, all the extra logs - which are naturally generated in the exact definition of the renormalized effective action $\sm_R$ - and are renormalization scale dependent, are not included in such formulation. These terms will break dilatation invariance and are not accounted for by the scaleless nonlocal action of Riegert type. 
Our work provides a more accurate view of such contributions, in a context in which DR is combined with dimensional reduction (DRed), in order to introduce a well-defined procedure for the derivation of the effective action at $d=4$.  

\subsection{DR with DRed}  
Finite renormalizations are a typical trait of renormalized theories with DR, but in the case of a curved manifold, several problems still need to be completely solved regarding the most appropriate way in which the $ d \to $ 4 limit should be performed. DR must be accompanied by an extension of the fields in the
variables of the extra dimensions, which is generally performed via a Kaluza Klein (KK) decomposition. A dimensional reduction procedure (DRed) is usually - at least implicitly - assumed, neglecting all the dependence on the variables of the extra dimensions.
In the KK decomposition this is equivalent to taking into account only the zero mode of the expansion on the extra dimensions.

The approach is quite similar to the previous analysis of the anomaly action in $ 4d $, where the term GB is introduced to satisfy the Wess-Zumino consistency condition from that action, and is not directly involved in the renormalization procedure. \\
The structure of the anomaly actions and that of the $ 4d $ EGB theory are discussed here in parallel, given the similarities. Several subtle points related to the presence of Weyl-invariant corrections coming from the
the choice of the extra dimensional metric and the regularization method, for actions of both types, are emphasized and carefully studied. Note that the $ 4d $ EGB theories are obtained from the anomaly action by eliminating the Weyl invariant terms that are generated by quantum corrections (the loop contributions).
These corrections originate from the only singular term in the action, $(1/\epsilon) V_{C^2}$ $(\epsilon=d-4)$, defined as the integral of the Weyl tensor squared $(C^2)$, introduced at 1-loop level in an anomaly action in order to regulate such contributions. We are going to denote them, in the next sections, as $\sm_f$ and $\tilde \sm_f$, depending on the regularization. \\
Note that $V_{C^2}$ and $V_E$ are analytic in $d$, and their formal expansions around $d=4$, at least for non singular metrics, are well defined. \\
At this point few remarks are in order. Conformal sectors induce single poles in the loop corrections, which are taken care of by the dimensional expansion of the counterterms.\\
The regularization of the pole in $1/\epsilon$, generated by the loop corrections, requires an expansion of the counterterms that can be performed in several ways, all differing by finite renormalizations. In our work, the Wess-Zumino action will represent only one of the possibile ways in which the subtractions are taken into account. 
Differently from the case of Minkowski space, where the issue does not arise, in a curved manifold extra dimensional components of the metric can be part of the pole residue. 
Similar issues emerge for $4d$ EGB theories, since the handling of the topological term follows the same pattern as for $V_{E/C^2}$.\\
While the singular limit of the coupling of a $4d$ EGB theories may look, at first sight, unmotivated, being the theory purely classical, it appears to be perfectly consistent, and generates a new class of Horndeski theories, classified as new forms of dilaton gravities. As we are going to show, a finite renormalization of $V_E$, that we will denote as $\hat V_E$, allows to remove the dilaton from the spectrum in such a theory.  \\
We are going to argue that both approaches - with or without the extra renormalization -  based on either of the two counterterms, identify effective actions which are useful in describing the effect of the anomaly at very different scales, covering either the UV or the IR, where by IR we refer to the scale (denoted by $f$) at which the conformal symmetry is broken by some extra sector. The UV effective action, on the other end, is appropriately expanded in terms of the dimensionless combination $R\Box^{-1}$ and captures the effect of the anomaly close to the Planck scale. In other words, both actions can be part of a unique renormalization group flow.

\section{The quantum effective action}
In this section we discuss the general structure of the quantum effective action, generated when a 
conformal sector is integrated out of the the partition function, and characterize its Weyl-invariant contributions.    
 
The backreaction of a conformal sector on the gravitational metric can be discussed via the partition function $\mathcal{Z}_B(g)$,
identified by the bare functional (in the Euclidean case)
\begin{equation}
\label{partition}
\mathcal{Z}_B(g)=\mathcal{N}\int D\chi e^{-S_0(g,\chi)},
\end{equation} 
where $\mathcal{N}$ is a normalization constant. We have denoted by $\chi$, just as example, a conformal scalar. \\
We will be deniting with ${-\mathcal{S}_B(g)}$ the 1-particle irreducible effective action,  defined as the log of the partition function 
\begin{equation}
\label{defg}
e^{-\mathcal{S}_B(g)}=\mathcal{Z}_B(g) \leftrightarrow \mathcal{S}_B(g)=-\log\mathcal{Z}_B(g). 
\end{equation}

In our case, quantum matter fields are assumed to be in a conformal phase. 
$\mathcal{S}_B(g)$ includes all the multiple insertions of the stress energy tensor 
  \bea
 \label{defT}
T^{\mu\nu}_{scalar}
&\equiv&\frac{2}{\sqrt{g}}\frac{\delta S_0}{\delta g_{\mu\nu}}\nonumber \\
&=&\nabla^\mu \chi \, \nabla^\nu\chi - \frac{1}{2} \, g^{\mu\nu}\,g^{\alpha\beta}\,\nabla_\alpha \chi \, \nabla_\beta \chi
+ \chi \bigg[g^{\mu\nu} \Box - \nabla^\mu\,\nabla^\nu + \frac{1}{2}\,g^{\mu\nu}\,R - R^{\mu\nu} \bigg]\, \chi^2, 
\eea

and diagrammatically corresponds to the expression

\begin{align}
\label{figgx}
\sm(g)=& \sum_n \quad\raisebox{-8.5ex}{{\includegraphics[width=0.19\linewidth]{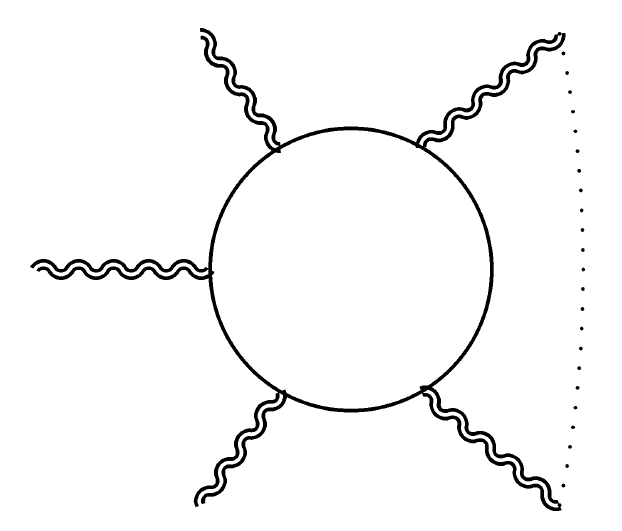}}} \,\scriptstyle \text{(n-point)}
\end{align}
which is expressed in  terms of stress energy tensor correlators $\langle T_1 T_2\ldots T_n\rangle$, with propagators and vertices that can be defined in any background, using \eqref{defT}.  The expansion can be constrained from the CWIs of the theory. \\
The simplest case that can be addressed is that of a flat background and, as shown in the figure above, can be computed by the ordinary Feynman expansion, order by order in $1/M_P^2$ in momentum space. The expansion accounts for the metric fluctuations $h_{\mu\nu}$, with  $g_{\mu\nu}=\delta_{\mu\nu} + h_{\mu\nu}$, generated by the virtual corrections due to the scalar field in the loops. In principle, one can use any background and of particular interest is the case of a De Sitter metric.  

In general, the contributions of such diagrams are divergent as $d\to 4$ and  need to be renormalized. In turn, this can be performed by the addition of the two counterterms $V_E$ and $V_{C^2}$,
causing the violation of the conformal symmetry in the effective action, as we will be discussing next. \\
The entire set of correlation functions of stress-energy tensors, to all orders in the fluctuations around certain metric background $\bar{g}$, is expressed in the form 
\begin{equation}
\label{exps2}
\sm(g)_B\equiv\sm(\bar{g})_B+\sum_{n=1}^\infty \frac{1}{2^n n!} \int d^d x_1\ldots d^d x_n \sqrt{g_1}\ldots \sqrt{g_n}\,\langle T^{\mu_1\nu_1}\ldots \,T^{\mu_n\nu_n}\rangle_{\bar{g} B}\delta g_{\mu_1\nu_1}(x_1)\ldots \delta g_{\mu_n\nu_n}(x_n),
\end{equation}
in terms of bare $(B)$ $nT$ correlators, with 
\begin{equation}
\label{exps1}
\langle T^{\mu_1\nu_1}(x_1)\ldots T^{\mu_n\nu_n}(x_n)\rangle_B \equiv\frac{2}{\sqrt{g_1}}\ldots \frac{2}{\sqrt{g_n}}\frac{\delta^n \sm_B(g)}{\delta g_{\mu_1\nu_1}(x_1)\delta g_{\mu_2\nu_2}(x_2)\ldots \delta g_{\mu_n\nu_n}(x_n)}, 
\end{equation}
where $\sqrt{g_1}\equiv \sqrt{|\textrm{det} \, g_{{\mu_1 \nu_1}}(x_1)} $ and so on. The renormalization of this functional expansion is rather involved in a general background, and can be best understood by borrowing the DR prescription around flat space. \\
In DR, the divergences appear as single poles, if we couple a conformal sector to gravity, and their renormalization, as already mentioned, is performed by expanding the counterterms around $d=4$. 
The two counterterms to be included are $V_E$ and $V_{C^2}$ that we will discuss next, giving a regularized effective action of the form 
\begin{equation}
\mathcal{Z}_R(g)_=\, \mathcal{N}\int D\Phi e^{-S_0(g,\Phi) + b' \frac{1}{\epsilon}V_E(g,d) + b \frac{1}{\epsilon}V_{C^2}(g,d)},
\end{equation} 
where $\mathcal{N}$ is a normalization constant. 
Here, $b$ and $b'$ count the number of massless fields involved in the loop corrections. \\
The role of the two counterterms is to remove the $1/\epsilon$ singularities present in the bare effective action $\sm_B$
\begin{equation}
\sm_B(g,d)=-\log\left(\int D\Phi e^{-S(\Phi,g)}\right) +\log\mathcal{N},
\end{equation}
and allow to define the regularized effective action in the form 
\begin{equation}
\label{rena}
\mathcal{S}_R(g,d)=\mathcal{S}_B(g,d) +  b' \frac{1}{\epsilon}V_E(g,d) + b \frac{1}{\epsilon}V_{C^2}(g,d).
\end{equation}
The expansion of the counterterms in the expression above is a critical step that needs a very close attention and must be checked by the choice of explicit metrics in the $d$ dimensional integrals $V_E$ and $V_{C^2}$. 
\section{The counterterms }
 As we are going to discuss next, all the issues concerning either the local or the nonlocal structure of the effective action 
 are related with the analysis of $V_E$ and $V_{C^2}$ and in our case they will be framed within DR. These two terms are defined in  terms of the Euler density $E$ and to the Weyl tensor squared $C^2$, respectively, by the expressions
\begin{align}
\label{ffr}
V_{C^2}(g, d)\equiv & \mu^{\varepsilon}\int\,d^dx\,\sqrt{-g}\, C^2, \notag \\
V_{E}(g,d)\equiv &\mu^{\varepsilon} \int\,d^dx\,\sqrt{-g}\,E , 
\end{align}
where $\mu$ is a renormalization scale while $\varepsilon=d-4$. The counterterm vertices will be simply obtained by multiple differentiations of the two integrals above. \\
We will omit $\mu$ from such counterterms in most of our analysis, just for simplicity, by setting $\mu\to 1$, and we will reinsert it into the final expression of the effective action, when we move from a  naive regularization of the action to a complete application of a standard DR/DRed procedure. It is important to remark that if we ignore the specific structure of the manifold of integration and assume the existence of some compactification for the extra $(d-4)$ dimensions, we have to face the problem of the presence of extra scales, beside $\mu$, in the effective action at $d=4$, with the generation of extra logarithms of the form $\log L \mu$. 
This point will be addressed rather carefully in the next sections. In DRed, such terms disappear as we take the $d\to 4$ limit, but this may not be general.  \\
The counterterms satisfy autonomous conformal Ward identities (CWIs) that can be solved in order to determine the trace anomaly contribution to each correlator  ($T, TT, TTT$ and so on), without the need of identifying the finite parts coming from the loops. This point has been discussed in \cite{Coriano:2021nvn, Bzowski:2018fql,Coriano:2018bbe,Coriano:2018bsy, Bzowski:2013sza,Bzowski:2015pba}. It should be clear that the anomaly part of the action that is responsible for the trace anomaly, obviously, does not account for the breaking of scale invariance that comes from a direct computation of the quantum corrections. This breaking is associated with the scale $\mu$. \\
 It has long been known that the inclusion of $V_E$ induces a finite renormalization of the effective action, since this term does not play any role in the cancelation of the singularities generated by the quantum corrections in the $d\to 4$ limit. In this respect, the use of the term "counterterm", is essentially a misnomer.
However, we will still use this expression when referring to it, just for simplicity.\\
  $V_E$ is introduced in order for the effective action to satisfy the Wess-Zumino (WZ) consistency condition. Indeed, at $d=4$, the integration of a conformal sector induces a renormalized effective action $\sm_R$ in \eqref{rena}, whose variation under an infinitesimal Weyl transformation of the metric 
 
\beq
\label{vars}
g_{\mu\nu}\to e^{2 \tau(x)} g_{\mu\nu},\qquad 
\delta_\tau g_{\mu\nu}= 2 \tau g_{\mu\nu} 
\eeq
 ($\delta/\delta_\tau=2 g_{\mu\nu}\delta/\delta g_{\mu\nu}$) is equal to the conformal anomaly. Notice that, as far as we stay in $d$ dimensions, the $\delta_\tau \mathcal{S}_R(d)$ 
 and the trace of the stress energy tensor generated by $\sm_R$ are identical, but  this is not the case as we take the $d\to 4$ limit of the renormalized effective action. 
 We are going to elaborate in more detail on this point in Section \eqref{cons1}.\\
If we integrate out a conformal matter sector at quantum level, the gravitational action is modified only by contributions up to second order in the Riemann tensor
\begin{equation}
 \delta_\tau \sm=\frac{1}{(4 \pi)^2}\int d^4 x \sqrt{g}\delta\, \tau(x)\left(c_1 R_{\mu\nu\rho\tau}R^{\mu\nu\rho\tau} + c_2 R_{\mu\nu}R^{\mu\nu} +c_3 R^2 + c_4\square R\right), 
\label{var}
\end{equation}
which are constrained by the Wess-Zumino consistency condition 
\begin{equation}
\label{WZ}
\left[\delta_{\tau_1},\delta_{\tau_2}\right]\sm_R=0,
\end{equation}
 and the coefficients $c_i$ have to satisfy the relation $c_1+c_2 +3 c_3=0$, allowing to re-express \eqref{var} in the form 
 \begin{equation}
 \label{anof}
 \delta_\tau\sm_R=\frac{1}{(4 \pi)^2} \int d^4 x \sqrt{g}\delta\tau(x) \mathcal{A}(x)
 \end{equation}
 where
\begin{equation}
 \label{vv}
 \mathcal{A}(x)=\left( a E + b C^2 + c\Box R\right) 
\end{equation}
is the conformal anomaly.
The coefficients $a,b,c$ are automatically fixed by the conformal sector that is integrated out, and the contributions $E$ and $C^2$ are both part of the variation of the renormalized effective action, generated by $V_E$ and $V_{C^2}$ contained in \eqref{rena}. 
Eq. \eqref{anof} is the usual expression of the conformal anomaly, generated by \eqref{rena}, with coefficients $a,b,c$ which are determined by the particle content of the theory: scalars, spin 1 vectors, and fermions $(n_s,n_V,n_f)$ that are integrated out at $d=4$. The last term ($\Box R$) is renormalization prescription dependent.\\ 
Since $\sm_B(g,d)$ is Weyl invariant, its Weyl variation according to  \eqref{vars} is zero 
and \eqref{anof} is entirely generated by the response of $\delta_\sigma V_E$ and $\delta_\sigma V_{C^2}$.
There are some subtle regularization issues, on which we will come back in the next sections, that need to be readdressed once we perform the $d\to 4$ limit 
of $\sm_R(g,d)$.\\ 
The first ambiguity comes from the definition of \eqref{anof} $C^2$,  the Weyl tensor squared in $d=4$ 
\begin{align}
( C^{(4)})^2&\equiv R_{\mu\nu\alpha\beta}R^{\mu\nu\alpha\beta}-2R_{\mu\nu}R^{\mu\nu}+\frac{1}{3}R^2, \label{fourd2}
\end{align}
which is generalized to $d$ dimension by the expression
\beq\label{Geometry1}
C^{(d) \alpha\beta\gamma\delta}C^{(d)}_{\alpha\beta\gamma\delta}
=
R^{\alpha\beta\gamma\delta}R_{\alpha\beta\gamma\delta} -\frac{4}{d-2}R^{\alpha\beta}R_{\alpha\beta}+\frac{2}{(d-2)(d-1)}R^2,
\eeq
where
\beq
C^{(d)}_{\alpha\beta\gamma\delta} = R_{\alpha\beta\gamma\delta} -
\frac{1}{d-2}( g_{\alpha\gamma} \, R_{\delta\beta} + g_{\alpha\delta} \, R_{\gamma\beta}
- g_{\beta\gamma} \, R_{\delta\alpha} - g_{\beta\delta} \, R_{\gamma\alpha} ) +
\frac{1}{(d-1)(d-2)} \, ( g_{\alpha\gamma} \, g_{\delta\beta} - g_{\alpha\delta} \, g_{\gamma\beta}) R.\, 
\eeq
The choice of one or the other version of $C^2$ affects the local part of the conformal anomaly functional, as discussed in the appendix.
The GB term, instead, is defined in $d$ dimensions in the form 
\beq
\label{GB1}
 E = R^2 - 4 R^{\mu \nu} R_{\mu \nu} + R^{\mu \nu \rho \sigma} R_{\mu \nu \rho \sigma}. 
\eeq
The relation between its $d$ dimensional expression and the $d=4$ is worked out, for an explicit metric choice, in \eqref{red}.

  \section{The regularized quantum effective action in DR} 
The relevant expression for the analysis of the effective action, here defined as $\sm_{R}$, starts from its definition in $d$ dimensions, as given in \eqref{rena}. The regularization of this action in DR is not unique. Similarly to the treatment of 
the perturbative expansion in flat space, one can always perform such procedure modulo the inclusion of finite contributions. 
These contributions, however, in flat space are generally harmless, and it is possible, at least in principle, to map one scheme to the other, although technical difficulties may be encountered.
On the other hand, in a curved background, even if we investigate the fluctuations around the flat limit of spacetime, we end up with actions characterized each by a different structure and field content.\\
We will be discussing two possible subtractions of the singularities present in the virtual corrections. One of them corresponds to the  usual (ordinary) DR approach, while the second one to the Wess-Zumino (WZ) subtraction, which is performed 
respect to a fiducial metric $\bar g$. A key relation which is important in order to identify the difference between the two schemes is given in Eq. \eqref{form1}.

The difference between the two methods is worked out in detail, and amounts to Weyl invariant terms.     

We implement DR on the counterterms, and use the analiticity of the two functionals $V$ respect to $d$, expanding their expressions around $d=4$, to obtain
\begin{equation} 
\label{expand1}
V_{E/C^2}(g, d)=\left( V_{E/C^2}(g, 4) + \varepsilon 
V_{E/C^2}'(g,4) +O(\varepsilon^2) \right),
\end{equation}
where only one background metric appears $(g)$.
The trace anomaly contribution is generated by this expansion. Different types of effective actions are generated, depending on the way we expand $V_{E/C^2}$.\\  
Establishing the true meaning of this formal expansion is a critical step that will require further elaborations. \\
Notice that \eqref{expand1} extends to curved space the usual DR approach of flat space. We recall that in DR - in flat space - once we turn to momentum space, we expand the residue of a  $1/\epsilon$ pole, starting from its value at $d=4$, plus finite terms. In general, the expansion will produce logarithms of the ratios of typical momenta of the diagrams and the renormalization scale $\mu$. Obviously, these finite logs are present in the computation of the loop corrections and break scale invariance. \\
The expansion plays an important role in our analysis of the renormalized action $\sm_R$, because, in principle, it can settle the controversy about the way the dilaton field shows up in the effective action, if linearly, quadratically or as a quartic power.  \\
The Weyl scalings in $d$ dimensions of $E$ and ${C^2}$ and their integrals $V_E$ and $V_{C^2}$ are crucial for investigating the structure of \eqref{expand1}.
We recall that $\sm_B$ in \eqref{rena} is only affected by single pole divergences in $1/\epsilon$, as we perform the $d\to 4$ limit. These can be isolated from its path integral expression that needs to be of the form  

\begin{equation}
\label{pone}
\sm_B(g,d)= \lim_{d\to 4}\left(\sm_f(d) -\frac{b}{\epsilon}V_{C^2}(g,4) - \frac{b'}{\epsilon}V_E(g,4)\right),
\end{equation}
where $\sm_f$ is finite. \eqref{pone} is justified by the fact that a conformal sector generates only singularities with  a single pole in all the correlators, and that these can be canceled just by the inclusion of $V_{C^2}$, accompanied by the evanescent term $V_E$.  

In $d$ dimensions, $\sm_B$ is finite and Weyl-invariant, but as we isolate the singular contributions from $\sm_B$ and perform the $d\to 4$ limit, the Weyl variation needs to be carefully redefined.  \\
 In Eq. \eqref{expand1} we are expanding the residue at the $1/\epsilon$ pole for any background metric $g$ and the $O(\epsilon^0)$ terms $V_{E/C^2}(4)$ need also to be treated with care. In  particular, the topological nature of $V_E(4)$, will guarantee that such term will not contribute to the (infinite) renormalization of the bare quantum action $\sm_B$ at $d=4$ , for being independent of any metric variation. Indeed, $V'_E$, and its variants, defined via the WZ part of $\sm_R$ in \eqref{rena}, that we will investigate below, correspond to finite renormalizations of all the correlation functions of stress energy tensors, generated by the functional expansion of $\sm_B$ or of $\sm_f$, once we take the $d\to 4$ limit. \\
 This behavior closely resembles the case of the chiral anomaly diagram, for example the AVV diagram, where a Chern Simons form can be introduced in order to preserve the vector Ward identities. Also in this case, the vertex does not need an infinite counterterm in order to regulate its two form factors which are divergent by power-counting. Since both the chiral anomaly and $ V_E $ are topological, this parallel is not surprising.
 
The procedure of renormalization can then be summarised by the expression 
\beqa
\label{sum}
\sm_R(d)&=&\Big(\sm_f(d)  -\frac{1}{\epsilon}V_{C^2}(g,4) - \frac{1}{\epsilon}V_E(g,4)\Big) +
\frac{1}{\varepsilon}\left(  V_{E}(g,4) + \varepsilon 
 V_{E}'(g,4) +O(\varepsilon^2) \right) \nn
&& \qquad + \frac{1}{\varepsilon}\left(  V_{C^2}(g,4) + \varepsilon 
 V_{C^2}'(g,4) +O(\varepsilon^2) \right), 
 \eeqa
where we have singled out, in the first bracket, the finite action $\sm_f$.
After reinserting the multiplicities $b$ and $b'$, we obtain
\begin{equation}
\label{sf1}
\sm_f(4)=\lim_{d\to 4}\Big(\sm_B(d)  +\frac{b}{\epsilon}V_{C^2}(g,4) +\frac{b'}{\epsilon}V_E(g,4)\Big)
\end{equation}
as evident from \eqref{pone}. Notice that $\sm_f(4)$ is invariant under a Weyl variation since $\sm_B(d)$ is Weyl invariant 
as well as $V_{C^2}(4)$ and $V_{E}(4)$.

In the expression of $\sm_R$ derived above, there is a cancellation between the $1/\epsilon$ contribution coming from $\sm_B$ (first bracket in \eqref{sum}) and those derived from the expansion of the counterterms $V_{E/C^2}$ (second and third bracket). 
The identification of such terms is quite involved, due to the need of computing propagators and vertices in a curved background. In few cases, they can be performed in DR using coordinate space methods. In the De Sitter case, for instance, such computations can be performed quite efficiently, especially for 1-point functions, such as for the $\langle T_{\mu\nu}\rangle$, using a regularization by point-splitting or by other techniques.   

After the cancellation of the singular terms in  \eqref{sum}, we are left with the renormalized effective action
\begin{equation}
\label{simp}
\sm_R\equiv \sm_R(4)=\sm_f (4)  + V'_E(g,4) + V'_{C^2}(g,4) 
\end{equation}
whose explicit structure will depend on the way this procedure will be 
implemented, by the choice of an explicit metric. 
Therefore, we can summarize the procedure in the $d\to 4$ limit by the expression 
\begin{equation}
\label{ren1}
\sm_R(4)=\lim_{d\to 4}\left(\sm_B(g,d) +\frac{b'}{\epsilon}V_E(g,d) + \frac{b}{\epsilon}V_{C^2}(g,d)\right)=\sm_f(4) +b' V'_E(\bar g,\phi, 4) + b V'_{C^2}(\bar g,\phi, 4),
\end{equation}
with 
\beq
\label{ps}
V'_{E/C^2}(\bar g,\phi,4)= \lim_{d\to 4}\left(\frac{1}{\epsilon}\left(V_{E/C^2}(g, d)- V_{E/C^2}(\bar g, 4)\right)\right),
\eeq
and the finite contribution coming from the loops contained in $\sm_f$
\beq
\sm_f(4) = \lim_{d\to 4}\left(\sm_B(d) +\frac{b'}{\epsilon}V_E(g,4) + \frac{b}{\epsilon}V_{C^2}(g,4)\right).
\eeq
The anomaly action generated by this regularization can then be defined in the form 
\beq 
\label{SA}
\sm_A=b' \,V'_{E}(\bar g,\phi,4) + b \,V'_{C^2}(\bar g,\phi,4),
\eeq
with $\sm_f(4)$ being the Weyl invariant part of $\sm_R(4)$. This part remains unaccounted for by $\sm_A$, unless we perform an explicit computation of the loop corrections. \\
Eq. \eqref{SA} is an important result that shows how the anomaly action can be identified just from the dimensional derivative of the two counterterms. We can check directly that    $V'_E$, for instance, reproduces the Euler part of the anomaly, using \eqref{ps}. We are going to discuss this point in more detail in the next sections.\\
 Notice that the subtraction term $V_E(g, 4)$ can be modified by replacing $g$ with $\bar g$, introducing a subtraction $V_E(\bar{g},4)$, where $\bar{g}$ is a fiducial metric. The two subtractions, as we are going to show, differ by Weyl invariant terms and determine different anomaly actions. 

Eqs  \eqref{SA} and \eqref{ps} are still quite formal, since they clearly depend on how we extend the 4-dimensional metric to $d$ dimensions and then reduce it to four.
We are going to provide explicit examples of the expansion above, which is motivated by the fact that $V_E$ and $V_{C^2}$ are both analytic in $d$ but are not uniquely defined in the DR procedure. Notice that both $V'_E$ and $V'_{C^2}$ take the form of local actions only if we extract a conformal factor from the metric, by introducing a conformal decomposition with respect to a fiducial metric $\bar{g}$. Notice that it is also possible to perform finite renormalizations of $E$ within DR, giving, as a result, anomaly actions which are quite different. For instance, as we are going to see, dilaton gravities of the form 
$4d-$EGB can be rewritten in the form of nonlocal actions.

 \section{Conformal decompositions and boundary terms}  
In this section we describe the behaviour of $V_{E}$ and $V_{C^2}$ under Weyl transformations, that will be essential in order to identify their dimensional expansions in $\epsilon$ around $d=4$. We assume that the Greek indices run from $1$ to $d$, unless otherwise specified. \\
This discussion is essential in order to underline the difference between the treatment of these counterterms in DR. 
Using the results summarized in the appendix,  the  GB density under a Weyl rescaling will change into the form
\beq\label{form1}
\rg E=\sqrt{\bar g} e^{(d-4)\phi}\biggl \{ \bar E+(d-3)\bar\nabla_\mu \bar J^\mu(\bar{g},\phi) +(d-3)(d-4)\bar  K(\bar{g},\phi)  \biggl \},
\eeq
where we have defined
\beq \label{GBexJ}
\bar J^\mu(\bar{g},\phi)=8\bar R^{\mu\nu}\bar\nabla_\nu\phi-4\bar R\bar \nabla^\mu\phi+4(d-2)(\bar\nabla^\mu\phi\bar \Box \phi-\bar \nabla^\mu\bar\nabla^\nu\phi\bar \nabla_\nu\phi+\bar\nabla^\mu\phi\bar\nabla_\lambda\phi\bar\nabla^\lambda\phi),
\eeq
\beq \label{GBexK}
 \bar K(\bar{g},\phi)=4\bar R^{\mu\nu}\bar\nabla_\mu\phi\bar\nabla_\nu\phi-2\bar R\bar\nabla_\lambda\phi\bar\nabla^\lambda\phi+4(d-2)\bar\Box\phi\bar\nabla_\lambda\phi\bar\nabla^\lambda\phi+(d-1)(d-2)(\bnabla_\lambda \phi\bnabla^\lambda \phi)^2.
\eeq
One important implication of the Weyl scaling expression above is the relation
\begin{equation} 
\label{ep2}
\frac{\delta}{\delta \phi}\int d^d y \sqrt{-g} E(y)=\epsilon \sqrt{g}E(x)
\end{equation} 
that can be derived in two ways, either by \eqref{form1}, as specified above or, more simply, by a metric variation. In the latter case one gets
\beq
\label{ep3}
 \dfun{}{g_{\mu \nu}} \int d^d x \rg E_4 = \rg \lt \frac{1}{2} g^{\mu \nu} E_4 - 2 R^{\mu \alpha \beta \gamma}R^\nu_{\alpha \beta \gamma} + 4 R^{\mu \alpha} R^\nu_{\ \alpha} + 4 R^{\mu \alpha \nu \beta} R_{\alpha \beta} - 2 R R^{\mu \nu} \rt  \eeq
and \eqref{ep2} follows if we contract with $2 g^{\mu\nu}$ both sides
\beq
\label{epx}
2 g_{\mu\nu}\frac{\delta}{\delta g_{\mu\nu}}\int d^d y \sqrt{-g} E(y)=\epsilon \sqrt{g}E(x).
\end{equation} 
This relation is true if we neglect boundary terms.  A direct computation, that accounts also for such terms, gives 
\beqa
2 g_{\mu\nu}(y)\frac{\delta}{\delta g_{\mu\nu}(y)}\int d^d x \sqrt{g}E &=& \epsilon \sqrt{g} E(y) -4 (d-3) \int d^d x \sqrt{g}
\nabla_\mu\left( R \nabla^\mu \delta_{x y}-2 R^{\mu\nu}\nabla_\nu \delta_{xy}\right)\nn
&=&\epsilon \sqrt{g} E(y) + 8(d-3) \int d^d x \sqrt{g}\nabla_\mu\nabla_\nu\left(G_{\mu\nu}(x)\delta^d_{x y}\right)
\eeqa
where $\delta_{x y}\equiv \delta^d(x-y)$ and with $G_{\mu\nu}$ denoting the Einstein tensor. The boundary term simplifies drastically, since the Einstein tensor is covariantly conserved and $G_{\mu\nu}$ can be pulled out of the integral, leaving the action of the covariant derivative only on the delta function
  \beq
  \int d^d x \sqrt{g}\nabla_\mu\nabla_\nu\left(G_{\mu\nu}(x)\delta^d (x-y)\right)=G_{\mu\nu}(y) \int d^d x \sqrt{g}\nabla_\mu\nabla_\nu \delta^d(x-y).
  \eeq
This contribution vanishes, since on the rhs of the equation above, the point $y$ is inside the region of integration, while we evaluate the integrand at an arbitrary distant boundary.   
Then we can re-express the relations above as 
\beq
2 g_{\mu\nu}\frac{\delta}{\delta g_{\mu\nu}}V_E(g,d)=\epsilon \sqrt{g} E + \textrm{boundary}
\eeq 
and there are no corrections of higher order in $\epsilon$. Obviously, this result is still affected by the contributions coming from the extra dimensions, and turns into a genuine $4d$ relation only in the presence of an explicit metric. For instance, as we are going to discuss in the next sections, if we choose a metric with a $4d$ dilaton, factorized overall as in \eqref{weylr}, then $E$ turns into a $4d$ density, if the extra dimensional metric is assumed to be flat, as indicated in \eqref{red}. \\
 Since $V_E(g,d) $ is a functional in which the dependence on the fiducial metric $\bar{g}$ is always accompanied by $e^{2 \phi}$, it respects the local $(\sigma)$ symmetry 
\beq
\label{sigma}
\bar{g}_{\mu\nu}\to \bar{g}_{\mu\nu}e^{2 \sigma} \qquad \phi\to \phi -\sigma,
\eeq
and the differentiation with respect to the conformal factor $\phi$ is equivalent to a metric differentiation plus a trace, at least in this case, giving
\beq
\label{bd}
 \frac{\delta}{\delta \phi}V_E(g,d)=\epsilon \sqrt{g} E + \textrm{boundary}.
\eeq 

This equivalence is broken in the presence of a regularization, as we are going to discuss in more detail in the next sections.
In the case of \eqref{bd}, the boundary can be derived from the rescaling relation \eqref{form1}. We obtain
\beqa
\label{mxx}
\frac{\delta}{\delta \phi}V_E(g,d)&=&\epsilon \sqrt{g}E + 
\int d^d x \sqrt{\bar g}e^{\epsilon \phi} \frac{\delta}{\delta\phi}\left(\bar E +\bar \nabla_\mu J^\mu(d-3) +(d-3)\epsilon K(\bar g,\phi)\right)\nonumber \\
&=&\epsilon \sqrt{g}E +(d-3) \int d^d x \sqrt{\bar g}\bar\nabla_\mu\left[  e^{(d-4)\phi}\left(\frac{\delta J^\mu }{\delta \phi} + 4(d-4)(d-2)\bar\nabla_\mu\phi\nabla_\nu\phi\nabla^\nu\delta_{xy}\right)\right].\nn
\eeqa 

A similar approach can be extended to $V_{C^2}$. Using

 \bea \dfun{}{g_{\mu \nu}} \int d^d x \rg \ C_{\alpha \beta \gamma \delta}^2 = \rg \Big(
\f g^{\mu \nu} C_{\alpha \beta \gamma \delta}^2 - 2 R^{\mu \alpha \beta \gamma} R^\nu_{\ \alpha \beta \gamma} + 4 R^{\mu \alpha} R^\nu_{\ \alpha} \nn - 4 \frac{d-4}{d-2}R^{\mu \alpha \nu \beta}R_{\alpha \beta} - \frac{4}{(d-2)(d-1)} RR^{\mu \nu} - 4 \frac{d-3}{d-2} \square R^{\mu \nu} \nn + 2 g^{\mu\nu} \frac{d-3}{(d-2)(d-1)} \square R + 2 \frac{d-3}{d-1} \nabla^\mu \nabla^\nu R \Big),  \eea 
and after a direct computation, one gets

\beq
\label{limit}
2 g_{\mu\nu}\dfun{}{g_{\mu \nu}} \int d^d x \rg C^2=\frac{\delta}{\delta \phi}\int d^d x \rg C^2=\epsilon \sqrt{g} C^2.
\eeq

Alternatively, by using the scaling relations above, one derives the condition $\sqrt{g}C^2=e^{\epsilon \phi}\sqrt{\bar{g}}{\bar C}^2$
and hence the equivalent relation 
 \beqa
 \frac{\delta}{\delta \phi}\int d^d x \sqrt{{g}}  {C}^2&=& \frac{\delta}{\delta \phi}\int d^d x e^{\epsilon \phi}\sqrt{\bar{g}}  \left(\bar{C}\right)^2\nn
 &=&\epsilon \sqrt{g}{C}^2.
 \label{xxs}
\eeqa

Notice that at $d=4$ this relation trivially becomes 
\beq
\label{xz1}
2 g_{\mu\nu}\dfun{}{g_{\mu \nu}} \int d^4 x \rg C^2=\frac{\delta}{\delta \phi}\int d^4 x \rg C^2=0
\eeq 
with a similar one for $V_E(g,4)$ 
\beq
\label{xz2}
2 g_{\mu\nu}\dfun{}{g_{\mu \nu}} \int d^4 x \rg E=\frac{\delta}{\delta \phi} \int d^4 x \rg E=0,
\eeq  
which is obvious, since this term is topological at $d=4$. 
As already mentioned, some relations concerning the variations of $C^2$ in $d=4$ and in general $d$ dimensions have been collected, for convenience in the  appendix.

\section{ The Wess-Zumino action versus the DR action} 
The  WZ effective action provides a regularization of the quantum effective action $\sm_R$ that differs from standard DR approach discussed above, by Weyl-invariant terms.\\
 Some ambiguities in the derivation of this part of the renormalized action can be noticed quite immediately. To illustrate this point, let's consider the conformal decomposition 

\beq
\label{mar}
g_{\mu\nu}=e^{2\phi(x)}\bar{g}_{\mu\nu} \qquad \bar{g}_{\mu\nu}= e^{-2\phi} g_{\mu\nu},
\eeq
expressed in terms of a fiducial metric $\bar{g}$ and a conformal factor $e^{2\phi(x)}$.
We recall that the regularization of the $V_E$ and $V_{C^2}$ terms 
 can be defined by a subtraction procedure of the form 
\beq
\label{WZ}
\hat{V}'_{E/C^2}(g,\phi)\equiv \lim_{d\to 4}\frac{1}{d-4} \left(V_{E/C^2}(\bar{g} e^{2 \phi},d) -V_{E/C^2}(\bar{g},d)\right).
\eeq

This specific definition of the counterterms, expanded in their dependence around $d=4$ and expressed in terms of the full metric and of the fiducial metric, as clear from \eqref{WZ}, is commonly used in the derivation of the  Wess-Zumino form (WZ) of the anomaly action
\beq
\sm_{WZ}= \hat{V}'_E(\bar g,\phi)+ \hat{V}'_{C^2}(\bar g,\phi).
\eeq

$\sm_R$ inherits a different decomposition, as one can figure out by going over the renormalization procedure discussed above. To get some insight into the derivation of $\sm_{WZ}$, similarly to \eqref{rena}, we reconsider the renormalized action, with the two counterterms expanded in a different form 
\beqa
\label{sumx}
\sm_R(d)&=&\Big( \sm_B(g, d) + \frac{1}{\epsilon} V_E(g,d) + \frac{1}{\epsilon}V_{C^2}(g,d) \Big)\nn\\
&=&  \sm_B(g,d) +\frac{1}{\epsilon}\left(V_E(\bar g,d)+\epsilon \hat V'_E(\bar g , \phi)\right) 
+\frac{1}{\epsilon}\left(V_{C^2}( \bar g,d)+\epsilon \hat V'_{C^2}(\bar g , \phi)\right),
\eeqa
where the expansion of the counterterms  is given by the relation \eqref{WZ}. Reorganizing the singular terms in order to remove  the singularity of  $\sm_B(d)$, 
we obtain
\beqa
\sm_R(d)&=&\Big( \sm_B(g,d)  +\frac{1}{\epsilon}V_{C^2}(\bar g,d) +\frac{1}{\epsilon}V_E(\bar g,d)\Big) +
\frac{1}{\epsilon} \left( V_{E}(g,d) 
- V_{E}(\bar g,d)  \right) + 
\frac{1}{\epsilon} \left( V_{C^2}(g,d) 
-V_{C^2}(\bar g,d)  \right), \nn
\eeqa
that we can rewrite in the form 
\beqa
\label{sumxp}
\sm_R(4)&=&\tilde\sm_f(4) +\sm_{WZ},
\eeqa

with 
\beq
\label{sf2}
\tilde\sm_f(4)=\lim_{d\to 4} \Big(\sm_B(d)  +\frac{1}{\epsilon}V_{C^2}(\bar g,d) +\frac{1}{\epsilon}V_E(\bar g,d)\Big).
\eeq
 As we are going to see next, the DR and WZ procedures differ by Weyl-invariant terms. These are taken into account in $\sm_f(4)$ \eqref{sf1} and $\tilde \sm_f(4)$ in two different ways. To determine all these contributions explicitly, we will be needing a careful implementation of DR, accompanied by a DRed procedure in order to perform the limit $d\to 4$. We pause for a comment.
 
The WZ action defined above, as already mentioned, is widely used in the context of the derivation of the anomaly actions in quantum gravity \cite{Coriano:2013nja,Coriano:2013xua}. In general, in these analysis, the attention goes either to the WZ part of $\sm_R$, denoted as $\sm_{WZ}$, or to $\sm_A$ defined in \eqref{SA}, and their completions in $\sm_R(4)$ are clearly different. These completions are $\tilde \sm_f (4)$ and $\sm_f(4)$, depending on whether we use either a standard DR or a WZ regularization of $\sm_R$.

\section{The anomaly-induced parts $\sm_A$ and $\sm_{WZ}$ in a generic regularization}
\label{cons1}
To get some additional insight into the contributions in \eqref{WZ} we proceed with a rescaling of the metric defining $V_E$, isolating the conformal factor. There are several subtleties related to the way we perform the limit in DR in order to expand the counterterms $V_E$ and $V_{C^2}$. We proceed with a regularization that we call "generic", since it is not based on an explicit metric parameterization. In a  follow-up section we are going to investigate the limitations of such method in more detail.

We will be using 
\eqref{form1} where all the fields are contracted on a $d$-dimensional spacetime manifold.
$V_E$ can be expanded in $\epsilon$ in the form 
\beq
\label{red1}
V_E(g,d)=\int d^d x \sqrt{\bar{g}}\left(\bar E + \bar\nabla_\mu \bar J^\mu \right) +
\epsilon\int d^d x \sqrt{\bar g} \phi\left( \bar E + \bar\nabla_\mu \bar J^\mu \right) 
+\epsilon \int d^d x \sqrt{\bar g} K +O(\epsilon^2).
\eeq
Notice that in the expression above, $O(\epsilon^2)$ contributions are also contained in the measure of integration, since this is still performed in $d$ dimensions. 
In a naive regularization, the limiting value of $V_E(g,d)$, denoted as $V_E(\bar g, 4)$, can be computed from 
\eqref{red1} by sending $\phi\to 0$ and setting directly $d=4$ in the integration measure 
\beq
\label{red2}
 V_E(\bar g,4)=\int d^4 x \sqrt{\bar{g}}\left(\bar E + \bar\nabla_\mu \bar J^\mu \right) =\int d^4 x \sqrt{\bar{g}}\bar E.
\eeq

We have dropped, in the last relation, a boundary term. Notice that, with the assumptions above, 
\beq
V_E(g,4)=V_E(\bar g,4),
\eeq
since all the dependence on the conformal factor $\phi$ disappears from both expressions. 
We need to be careful with such expressions since their 4-dimensional form may include extra contributions coming from the integration over the extra dimensions, inducing extra components of the metric in the reduced action. In general, the reduction to the base manifold, does not correspond to a real compactification, where extra Kaluza-Klein modes need to be taken into account. \\
Implicit, in most of the literature on the subject, is often the use of DRed without mentioning it. Here we  intend to address it more rigorously, by keeping track of extra contributions and extra cutoffs that are generated by the procedure. Note that \eqref{red2} is defined unambigously, being expressed in terms of the fiducial metric at $d=4$. $V_E(\bar g,d)$ can be defined analogously, by sending $\phi\to 0$ in \eqref{red1} 
\beq
 V_E(\bar g,d)=\int d^d x \sqrt{\bar{g}}\left(\bar E + \bar\nabla_\mu \bar J^\mu \right) =\int d^d x \sqrt{\bar{g}}\bar E.
\eeq
 The complete contribution to $\sm_A$ coming from the $V_E$ counterterm is then given by 
\beq
\label{like}
V'_E\equiv\lim_{d\to 4}\left( \frac{1}{\epsilon}\left( V_E(g,d)-V_E(g,4)\right)\right)=
\frac{\partial}{\partial d}V_E(\bar g, d)\mid_{d=4} +
\int d^4 x \sqrt{\bar g} \phi\left( \bar E + \bar\nabla_M \bar J^M \right) 
+\int d^4 x \sqrt{\bar g} K,
\eeq	
where the first terms on the rhs is due to the difference 
\beq
\frac{\partial}{\partial d}V_E(\bar g, d)\mid_{d=4}=\lim_{d\to 4}\left(\frac{1}{\epsilon}\left( \int d^d x \sqrt{\bar{g}}\bar E - \int d^4 x \sqrt{\bar{g}}\bar E\right)\right)
\label{difE}
\eeq

which differs respect to the renormalization used in the definition of the WZ effective action by the equation
\beq
\label{wzp}
\hat{V}'_E(\bar g,\phi)\equiv\lim_{d\to 4}\left(\frac{1}{\epsilon}\left(V_E(g,d)-V_E(\bar g, d)\right)\right)=V'_E- \frac{\partial}{\partial d}V_E(\bar g, d)\mid_{d=4},
\eeq
that gives
\beq
\label{wzp2}
\hat{V}'_E(\bar g,\phi)=\int d^d x \sqrt{\bar g} \phi\left( \bar E + \bar\nabla_\mu \bar J^\mu \right) 
+\int d^d x \sqrt{\bar g} K.
\eeq

A similar analysis, in the case of the quantum effective action, can be performed for $V_{C^2}$ with

\beq
\label{cc}
V_{C^2}(g,d)=\int d^d x \sqrt{\bar g} e^{\epsilon \phi} \bar C^2 =\int d^d x \sqrt{\bar g}  \bar C^2 
+\epsilon \int d^d x \sqrt{\bar g} \phi \bar C^2, 
\eeq
giving
\beq
 V_{C^2}(\bar g,4)=\int d^4 x \sqrt{\bar g}\bar C^2 
\eeq
where, also in this case, 
\beq
 V_{C^2}(\bar g,4)=V_{C^2}(g,4).
\eeq
In DR, the contribution to the anomaly action is then given by  

\beq
V'_{C^2}\equiv \lim_{d\to 4}\left(\frac{1}{\epsilon}\left( V_{C^2}(g,d)-V_{C^2}( g,4)\right)\right)=
\frac{\partial}{\partial d}V_{C^2}(\bar g, d)\mid_{d=4} + \int d^d x \sqrt{\bar g} \phi \bar C^2,
\eeq
where 
\beq
\frac{\partial}{\partial d}V_{C^2}(\bar g, d)\mid_{d=4}= \lim_{d\to 4}\left(\frac{1}{\epsilon}\left( \int d^d x \sqrt{\bar{g}}\bar C^2 - \int d^4 x \sqrt{\bar{g}}\bar C^2\right)\right).
\label{difC}
\eeq
The contribution of the regulated counterterm to the WZ action is computed similarly to \eqref{wzp2}
\beq
\hat V'_{C^2}(\bar g,\phi)=  \lim_{d\to 4}\left(\frac{1}{\epsilon}\left( V_{C^2}(g,d)-V_{C^2}( \bar g,d)\right)\right)=\int d^d x \sqrt{\bar g} \phi \bar C^2.
\eeq
It is then clear that $\sm_A$ and $\sm_{WZ}$ 
differ by the  contributions  
\beq
\sm_A - \sm_{WZ} =\sm_{A/WZ}  \eeq
where
\beq
\sm_{A/WZ}\equiv \frac{\partial}{\partial d}V_E(\bar g, d)\mid_{d=4} + 
\frac{\partial}{\partial d}V_{C^2}(\bar g, d)\mid_{d=4}.
\eeq
which correspond to Weyl invariant terms, being only dependent on the fiducial metric $\bar{g}$, as defined by \eqref{difE} and \eqref{difC}.

The ambiguities in the definition of the Weyl-invariant terms is a natural result of the renormalization procedure, due to the prescriptions used in the regularization of the effective action. In general, the main difference between the different regularizations lays in the power of the dilaton field. In $\sm_A $ and $\sm_{WZ}$ the dependence on $\phi$ is quartic, but the inclusion of a finite renormalizaton of the topological density makes it quadratic. This point will be addressed in \secref{mod}.

By the term "anomaly induced actions" we will refer to both the $\sm_A $ and $\sm_{WZ}$ contributions, which do not include the 
 finite term $\sm_f$ or $\tilde{S}_f$  coming from the quantum corrections. The conformal backreaction is associated with $\sm_R$, and includes also $\sm_f$ and $\tilde{S}_f$, as defined in \eqref{sf1}. Their invariance under $\phi$-variations ($\sim \delta_\phi$), does not imply that they will not contribute to the scale anomaly. We will come back shortly to this point, which is slightly involved.\\
 For the moment, we can summarize the relation between the two different regularizations of the complete effective action 
 $S_R$ by the equations
 \beq
 \sm_R=\sm_f + \sm_A
 \eeq
 and
 \beq
\sm_R=\tilde\sm_f + \sm_{WZ},
 \eeq
 where
 \beq
 \tilde\sm_f -\sm_{{f}}= \sm_A -\sm_{WZ}=\sm_{A/WZ}.
 \eeq
 
 To recover the $4d$ EGB theory we need to retain in the relation above only the regulated GB term, and the corresponding action will be simply given by 
 the two actions 
 \beqa
 \label{sxc}
 \sm_{EGB/1}&=&S_{EH} + V'_E(\bar{g},\phi) \nn
 \sm_{EGB/2}&=&S_{EH} + \hat V'_E(\bar{g},\phi), 
 \eeqa
 denoted as EGB/1 and EGB/2, where $\sm_{EH}$ is the Einstein-Hilbert action. The two actions, as we have, seen, differ by Weyl-invariant contributions generated by the regularization. As we are going to discuss next, there are several relevant points that need to be addressed concerning the explicit expressions of $V'_E(\bar{g},\phi)$ and  $\hat V'_E(\bar{g},\phi) $. 
The dilaton can be removed if we introduce a modification of the GB density away from $d=4$, extension already implemented in the past in the case of anomaly actions, for the treatment of this term.  \\
Therefore, in the case of $4d$ EGB theories, which share similarities with anomaly actions, the equations of motion still have to satisfy constraints which are similar to those generated by the Weyl variant counterms $V'$. These actions are simply defined by adding to the EH action, the $\hat V'_E$ term, as shown in \eqref{sxc}. This clarifies the origin of the constraints found for these theories in \cite{Hennigar:2020lsl} in their equations of motion.

 \subsection{Anomaly constraints} 

At this point, we are in condition to derive the anomaly constraints on $\mathcal{S}^{WZ}$ and on 
$\sm_A$, clarifying some of the intermediate steps in the derivations. We recall that the usual relation
\beq
\label{pp}
2 g_{\mu\nu}\frac{\delta}{\delta g_{\mu\nu}}=\frac{\delta}{\delta\phi},
\eeq
holds only if, as a precondition, a functional can be written in terms of the entire metric $g$. The identiy holds as far as we perform variations of $g$ and keep the fiducial metric $\bar{g}$ in \eqref{mar} fixed. \\ 
The usual derivation of the condition of Weyl invariance, is to perform a conformal decomposition of $g$ in terms of a fiducial metric, and verify that there is no dependence of the functional on $\phi$. In other words, Weyl  invariance is equivalent to the condition of independence from the conformal factor. However, the precondition clearly tells us that we should not perform any subtraction of $\bar{g}$ terms on this functional. For this reason, the precondition can be formulated as the request of invariance of the functional respect to the conformal decomposition in \eqref{mar},
\beq
\label{b1}
\delta_\sigma F=0 \qquad \textrm{with} \qquad g_{\mu\nu}\to g_{\mu\nu}e^{2 \sigma},\qquad  \phi\to \phi -\sigma. 
\eeq
This second constraint is clearly violated by the renormalization procedure, that forces us to identify a specific conformal decomposition, fixed by a scale $f$ appearing in the conformal factor. 

In this case, the functional cannot be rewritten as a function of the entire metric $g$.
It is then clear that once we view the $\bar{g}$ term in the WZ action as a subtraction, needed in order to perform a DR/DRed regularization of the quantum corrections in $\sm_R$, it is misleading to refer to the subtraction term $V_{E/C^2}(\bar g, d)$ as being "Weyl invariant" respect to $V_{E/C^2}(g, d)$, the "Weyl variant" one.\\
To summarize: once we perform  the subtraction of the pole terms using $V_{E/C^2}(\bar g, d)$ on $\sm_B$, a naive implementation of this variation on this term would obviously give zero
\beq
\delta_\phi V_{E/C^2}(\bar g, d)=0
\eeq
but the $\delta_\sigma$ symmetry is violated both in $\sm_A$ and in $\sm_{WZ}$. 
The breaking of both symmetries ($\delta_\sigma$ and $\delta_\phi$), now interpreted as 
a breaking of Weyl invariance, clarifies why there is also a breaking of scale invariance and not only a generation of the trace anomaly, in the complete effective action, after renormalization. \\
Weyl invariance, as mentioned, should be associated with the conformal symmetry of a certain functional, such that if we expand the parameter of the Weyl transformation, say $\sigma(x)$, around flat space, then this function is constrained to be at most quadratic in $x$. It would be expressed in terms of 15 parameters, paired with the generators of the conformal group, one of them being the dilatation. An anomaly induced action such that its $\delta_\phi$ variation equals the anomaly, and interpreted as a Weyl-variant functional, without any reference to the renormalization procedure, falls short from predicting the breaking of scale invariance. From this point of view, it is clear that $V_{E/C^2}(\bar g,d)$ is as essential as the remaining $\delta_\sigma$ invariant term $V_{E/C^2}(g,d)$ in the definition of the anomaly effective action. It is clear that, after renormalization, $\phi$ and $\bar{g}$ should be treated as independent fields and varied accordingly, since the 
$\phi$ variation alone is performed only for a fixed fiducial metric. This is required for the derivation of \eqref{pp}. \\
The explicit breaking of the combined $\delta_\sigma$ and $\delta_\phi$ variations, allows us to justify the presence of the other scales $(\mu,f)$, introduced by the renormalization procedure.  

\subsection{Implications for the functional differentiation}

Given a functional $F(g)$, that satisfies the constraint $\delta_\sigma F=0$, we use the relation $\bar{g}_{\mu\nu}=g_{\mu\nu}e^{-2 \phi}$, 
to derive the expressions
\beq
\label{rels}
\frac{\delta\bar{g}_{\mu\nu}(x) }{\delta g_{\alpha\beta}(y)}=\delta^{\alpha\beta}_{\mu\nu}e^{-2 \phi}\delta^d(x-y),
\qquad 
\frac{\delta F}{\delta g_{\alpha\beta}}= \frac{\delta F}{\delta \bar{g}_{\alpha \beta}}e^{-2 \phi},\qquad 
2 \bar g_{\alpha\beta}\frac{\del F}{\delta\bar g_{\alpha \beta} }=2  g_{\alpha\beta}\frac{\del F}{\delta g_{\alpha \beta} }.
\eeq
The regularization, by separating the dependence on the two components of $g$, leads to 
identities between the functional variations w.r.t. $\phi$ and $\bar{g}$ that, as we are going to show, are related by 
the anomaly.

For instance, from $\sm^{WZ}$, the contribution from the Euler density satisfies the relation
\beq
2 g_{\mu\nu}\frac{\delta \hat V'_E(\bar g,\phi)}{\delta g_{\mu\nu}}=
\frac{1}{\epsilon}\left(2 g_{\mu\nu}\frac{\delta V_E(g,d)}{\delta g_{\mu\nu}}-
2 g_{\mu\nu}\frac{\delta V_E(\bar g,d)}{\delta g_{\mu\nu}}\right).
\eeq
Now using 
\beq
2 g_{\mu\nu}\frac{\delta}{\delta g_{\mu\nu}}V_E(\bar g,d)=2\bar g_{\mu\nu}\frac{\delta}{\delta \bar g_{\mu\nu}}V_E(\bar g,d)=\epsilon\sqrt{\bar g}\bar E,
\eeq
together with \eqref{epx}, we obtain 
\beq
2g_{\mu\nu}\frac{\delta \hat V'_E}{\delta g_{\mu\nu}}=\sqrt{g}E -\sqrt{\bar g }\bar E.
\eeq
On the other end, we have, for any functional in which $\bar{g}$ and $\phi$ are paired, Eq. \eqref{pp} is satisfied and gives 

\beq
\label{ord}
\frac{\delta}{\delta \phi}V_E(g,d)=\epsilon \sqrt{g}E, 
\eeq
while 
\beq
 \frac{\delta}{\delta \phi}V_E(\bar g,d)=0,
\eeq
that combined together give
\beq
\label{equal}
\frac{\delta}{\delta \phi}\hat V'_E(\bar g,\phi) =\sqrt{g}E.
\eeq
Therefore, in the case of a $4d$ EGB theory, as well as for any anomaly action, one derives a constraint between the equation of motion of the dilaton and the trace of the stress energy tensor of the fiducial metric  
\beq
\label{break}
2g_{\mu\nu}\frac{\delta \hat V'_E(\bar g ,\phi)}{\delta g_{\mu\nu}}-\frac{\delta  \hat V'_E(\bar g ,\phi)}{\delta \phi}=-\sqrt{\bar g}\bar E.
\eeq

This relation shows that the Weyl variation and the trace/metric variation are not identical, if we perform a regularization.
A similar relation holds for $\hat V'_{C^2}$
\beq
\label{break1}
2g_{\mu\nu}\frac{\delta \hat V'_{C^2}(\bar g ,\phi)}{\delta g_{\mu\nu}}-\frac{\delta \hat V'_{C^2}}{\delta \phi}=-\sqrt{\bar g}\bar C^2.
\eeq

In summary, a WZ anomaly action, following the definitions above, will satisfy the anomaly condition 
\beq
\label{WZZ}
2g_{\mu\nu}\frac{\delta \mathcal{S}^{WZ}}{\delta g_{\mu\nu}}=b'\sqrt{g}E + b \sqrt{g} C^2 - 
\left(b'\sqrt{\bar g}\bar E + b \sqrt{\bar g} \bar C^2\right),
\eeq
and 
\beq
\label{WZZ1}
\frac{\delta \mathcal{S}^{WZ}}{\delta \phi}=b'\sqrt{g}E + b \sqrt{g} C^2 
\eeq
and, more generally, the condition 

\beq
\label{cons}
2g_{\mu\nu}\frac{\delta \mathcal{S}^{WZ}}{\delta g_{\mu\nu}}-\frac{\delta \mathcal{S}^{WZ}}{\delta \phi}= - 
\left(b'\sqrt{\bar g}\bar E + b \sqrt{\bar g} \bar C^2\right).
\eeq

It differs from the DR anomaly-induced action by terms 
\beq
\label{nv2}
\left(2g_{\mu\nu}\frac{\delta \mathcal{S}^{A}}{\delta g_{\mu\nu}}- 2g_{\mu\nu}\frac{\delta \mathcal{S}^{WZ}}{\delta g_{\mu\nu}}\right)= \frac{\partial}{\partial d} V_E(\bar g,d)\mid_{d=4} + \frac{\partial}{\partial d} V_{C^2}(\bar g,d)\mid_{d=4},
\eeq
that are Weyl invariant, since they do not depend on the conformal factor $\phi$. 

\subsection{$\sm_B(d)$ for $d\to 4$}
 $\sm_B(d)$ describes the finite quantum corrections that develop a singularity in the $d\to 4$ limit. Notice that this is a functional of the entire metric before the limit is taken, and it is Weyl-invariant. Therefore it satisfies the constraint
 \beq
 \label{ccd}
 2 g_{\mu\nu}\frac{\delta\sm_B }{\delta g_{\mu\nu}}=\frac{\delta \sm_B}{\delta\phi}=0.
 \eeq
 This relation implies that $\sm_B$ is only a functional of $\bar{g}$ since a Weyl variation does not change its functional expression
 \beq
 \sm_B(g,d)=\sm_B(\bar g). 
 \eeq
 
 Using \eqref{xz1} and \eqref{xz2}, that we rewrite in the form 
 \beq
2 g_{\mu\nu}\frac{\delta V_{E/C^2}(g,4) }{\delta g_{\mu\nu}}=\frac{\delta V_{E/C^2}(g,4)}{\delta\phi}=0,
\eeq

 one finds that $ \sm_f(d)$
  \beq
 \sm_f(d) =\lim_{d\to 4}\left(S_B(g,d) + b'\frac{1}{\epsilon}V_{E}(\bar g ,4) + b\frac{1}{\epsilon}V_{C^2}(\bar g ,4)\right),
 \eeq
  is only a functional of the fiducial metric, and its stress-energy tensor has a vanishing trace
 
 \beq
 \frac{\delta}{\delta \phi}\sm_f(\bar g,\phi, d)= 0.  
 \eeq
 
 This property continues to hold in the  $d\to 4$ limit and henceforth 
\beq
\sm_f(4)\equiv \sm_f(\bar g).
 \eeq
This implies that $\sm_f(4)$ does not contribute to the trace anomaly and we have consistently the constraints 

\beq
\label{course}
2 \bar g_{\mu\nu}\frac{\delta \sm_f(4)}{\delta \bar g_{\mu\nu}}=0, 
\qquad \frac{\delta \sm_f(4)}{\delta \phi}=0. 
\eeq
Notice, however, that this condition does not guarantee that $\sm_f(4)$ is also scale-invariant. Indeed, it is not. As we have mentioned, logs of the renormalization scale $\mu$ are present in this functional and in $\sm_R$ after renormalization, due to the breaking of the local shift symmetry \eqref{sigma}. According to this symmetry, $\phi(x)$ takes the role of a Nambu-Goldstone mode. The invariance of the functional under a local shift is indeed broken by the renormalization procedure. This observation clarifies why terms in the anomaly action that are Weyl invariant (i.e. they are terms that depend only on the fiducial metric $\bar{g}$) break the dilatation symmetry. 
In the WZ case, instead, the subtraction is defined as
\beq
\tilde\sm_f(d) =\lim_{d\to 4}\left(S_B(g,d) +b'\frac{1}{\epsilon}V_{E}(\bar g ,d)+ b\frac{1}{\epsilon}V_{E}(\bar g ,d)\right),
 \eeq
and in this case one easily derives the relation 
\beq
\frac{\delta\tilde\sm_f}{\delta \phi}=0, 
\eeq
since the subtractions $V_{E/C^2}(\bar g ,d)$ and the bare action $\sm_B$ do not depend on $\phi$. On the other hand, we have 
\beq
2 \bar{g}_{\mu\nu}\frac{\delta V_{E}(\bar g ,d)}{\delta \bar g_{\mu\nu}}=\sqrt{\bar g} \bar E,
\eeq
and similarly for $V_{C^2}(\bar g ,d)$, thereby obtaining

\beq
\label{dd2}
2 \bar{g}_{\mu\nu}\frac{\delta \tilde\sm_f(\bar g ,d)}{\delta \bar g_{\mu\nu}}=
b'\sqrt{\bar g}\bar E  +b\sqrt{\bar g}\bar C^2. 
\eeq

\subsection{The complete quantum action in DR}
It is clear that the conformal backreaction 
is associated with the entire renormalized effective action $\sm_R$, rather than with the anomaly-induced actions $\sm_A$ or the WZ action $S^{WZ}$. The difference between $\sm_R$ and the previous two actions is, again, given by Weyl invariant terms. 
We recall that, as far as we stay in $d$ dimensions, $S_{R}(g,d)$, defined by the sum of 
$\sm_B(g,d)$ and of the two counterterms $1/\epsilon V_E(g,d)$ and $1/\epsilon V_{C^2}$, under a Weyl variation behaves as
\beq
\frac{\delta}{\delta\phi}\sm_R(g,d)=\frac{\delta \sm_B(g,d)}{\delta\phi}  +\frac{\delta}{\delta\phi}\Bigg(\frac{1}{\epsilon}\left(b' V_E(g,d) + b V_{C^2}(g,d)\right) \Bigg).
\eeq
Then, using the invariance of $\sm_B$ \eqref{ccd}, we obtain
\beq
 \frac{\delta \sm_R(g,d)}{\delta\phi}=2 g_{\mu\nu}\frac{\delta \sm_R(g,d)}{\delta g_{\mu\nu}}=
 b'\sqrt{g}E + b\sqrt{g}C^2.
\eeq
This equation is modified by the separation into poles in $1/\epsilon$ plus finite terms in DR as follows 
\beq
\frac{\delta \sm_R(g,4)}{\delta\phi}=\frac{\delta \sm_f(g,4)}{\delta\phi} +\frac{\delta \sm_A(g,4)}{\delta\phi}.
\eeq
In this case both a $\phi$- variation and trace-metric variation coincide and give 

\beq
\label{1p}
\frac{\delta \sm_A(4)}{\delta\phi}= b'\sqrt{g}E + b\sqrt{g}C^2
\eeq 
and

\beq
\label{2p}
2 g_{\mu\nu}\frac{\delta \sm_A(4)}{\delta g_{\mu\nu}}=b'\sqrt{g}E + b \sqrt{g} C^2. 
\eeq 

The finite part of $\sm_R$, given by $\sm_f$, has no dependence on the conformal factor, and its stress energy tensor has zero trace. Therefore the anomaly, in this case, is all generated by $\sm_A$
\beq
\label{canc}
\frac{\delta\sm_R}{\delta{\phi}}=2g_{\mu\nu}\frac{\delta \sm_R}{\delta g_{\mu\nu}}= 
 b'\sqrt{g}E + b\sqrt{g}C^2.
\eeq
 
 If we perform the renormalization of $\sm_B$ using \eqref{sumx} and \eqref{sumxp}, 
 then \eqref{canc} is obviously still valid. However both $\tilde\sm_f$ and $\sm_{WZ}$  on the rhs of \eqref{sumxp} will contribute to the trace anomaly, with extra Weyl-invariant terms that carry opposite signs, as shown in \eqref{WZZ} and \eqref{dd2}, their sum reproducing again Eq. \eqref{canc}.
 
 \section{Effective actions in extra dimensional schemes with dimensional reduction}
 \label{EDDR}
 In this section we proceed with a different analysis of the extra dimensional (ED) contributions, which in the previous section have been generic, by considering a specific metric choice. We will factorize its $d$-dimensional expression into a general conformal factor times a fiducial metric, taking the topology of a direct product. The ED space will be, at the end, assumed to be flat, and we will impose the condition of dimensional reduction DRED on the 4-dimensional curvatures and fields. 
 In normal Kaluza-Klein (KK) compactifications, this is equivalent to taking the zero mode of the KK towers of the metric, which does not depend on the geometry of the $d-4$-dimensional outer space. This approach will introduce a universal cutoff $L$ on the size of the ED space, that will be accompanied by the renormalization scale $\mu$ in the logarithmic corrections to the anomaly actions, generating logarithms of the dimensionless variable $L\mu$. \\
 We start by analyzing the topological counterterm, by mentioning that 
from \eqref{form1}, using \eqref{GBexJ} and \eqref{GBexK}, we obtain the relation
\beq
\bar \nabla_\mu\phi \bar J^\mu-\bar K=4\bar R^{\mu\nu}(\bnabla_\mu\phi\bnabla_\nu\phi)-2\bar R\bar \square \phi+2(\bnabla_\lambda \phi \bnabla^\lambda \phi )^2+4\bar\Box\phi\bnabla_\lambda \phi \bnabla^\lambda \phi ,
\eeq
written as a four-dimensional expression, but that is also valid  in the embedding space. As far as we do not specify the extra dimensional metric in some way, the Greek indices may be used, with no confusion, 
to describe the invariants over the entire embedding space.
 
After an integration by parts, we get the final form of the counterterm contribution, up to $O(\epsilon)$ terms, given  by
\bea \label{WG in generic case}
\frac{1}{\epsilon}V_E(g,d) &=& \frac{1}{\epsilon}\int d^dx \rgb \ \bar E+ \int d^dx \rgb\ \Big[\phi\bar E-(4\bar G^{\mu\nu}(\bar\nabla_\mu\phi\bar\nabla_\nu\phi) \nn && +2(\bnabla_\lambda \phi \bnabla^\lambda \phi )^2 +4\bar\Box\phi \bnabla_\lambda \phi \bnabla^\lambda \phi ) \Big].
\eea

This result holds in $d$ spacetime dimensions. 

If we intend to 
take the $d\to 4$ limit with more rigour, then we need to be more specific about the choice of the metric. In our case, for definiteness, we will consider a manifold of the form $\mathcal{M}_4\times \mathcal{M}_e$, split into a 4- and $(d-4)$-dimensional part.  
We will denote the $d$ dimensional indices as $M, N$, saving the Greek indices for the 4-dimensional part.  The $d$ dimensional metric is decomposed in the Weyl gauge and split into the direct sum of the metrics of the two submanifolds. ${}_e g_{m n }$ is the extra dimensional metric used for the regularization of the integral and $_4\tilde{g}_{\mu\nu}$ is its 4-dimensional part. The extra coordinates will be denoted as $y$. For example, we choose  
\beq
\apd g_{MN}(x,y)  =  e^{2 \phi(x)} \begin{bmatrix}  {}_4\tilde g_{\mu\nu}(x) &0\\
0 &  \ape  g_{mn}(y)
\end{bmatrix}  = e^{2 \phi(x)}\apd \bar g_{MN} .
\label{weylr}
\eeq
$_d\bar{g}_{\mu\nu}$ is the $d$-dimensional fiducial metric, from which we have extracted a conformal factor $\phi$, with the indices decomposed as $M=(\mu,m), \, (N=\nu,n)\ldots$, and so on.   
The original scaling relation \eqref{form1} in $d$ dimensions can be expressed, in this spacetime manifold, in the form
\beq \label{iter1a}
\int d^4x d^{d-4}y \rgd \ \apd E =  \int d^4x d^{d-4}y  \rgdb\ e^{(d-4)\phi} \lt \apd \bar E+(d-3)\bar\nabla_M \apd \bar J^M+(d-3)(d-4) \apd \bar K  \rt ,\eeq
where barred terms, including the covariant derivatives, are relative to the fiducial metric $\apd \bar g_{MN}$.
Since, with our choice, the dilaton does not appear in the extra dimensional part of the metric, the two blocks that make up the full $d$-dimensional metric $\apd \bar g_{MN}$ are only dependent on the coordinates of the submanifold which they belong to. The same is true for every curvature tensor, hence we have that ${}_4\tilde R_{\mu \nu \rho \sigma}$, the Riemann tensor of the base space, depends only on the $x$-coordinates, while ${}_e R_{abcd}$ depends only on the $y$-coordinates. Moreover, the connection has no mixed terms, hence the squared curvature tensors are decoupled
\bea \apd \bar R^{ABCD} \apd \bar R_{ABCD}&=& {}_4 \tilde R^{\mu \nu \rho \sigma} {}_4 \tilde  R_{\mu \nu \rho \sigma} + \ape R^{abcd} \ape R_{abcd}, \nn
\apd \bar R^{AB} \apd \bar R_{AB}&=& {}_4 \tilde R^{\mu \nu} {}_4 \tilde R_{\mu \nu} + \ape R^{ab} \ape R_{ab}, \nn
\apd \bar R^2 &=& {}_4 \tilde R^2 + \ape R^2 + 2 {}_4 \tilde R \ape R . \label{squaredcurvaturefact} \eea
The Gauss-Bonnet density, for example, 
becomes 
\beq 
\label{red}
\apd \bar E = {}_4 \tilde E + \ape E + 2 {}_4 \tilde R  \ape R . \eeq
From the definitions \eqref{GBexJ} and \eqref{GBexK}, since $\phi(x)$ is a function only of the coordinates of the 4-dimensional subspace, we obtain
\bea 
\apd \bar J^\mu &=& {}_4 \tilde J^\mu - 4\ape R\ \tilde \nabla^\mu \phi , \qquad \ape \bar J^m = 0 ,\nn
\apd \bar K &=& {}_4 \tilde K - 2\ape R\ \tilde \nabla^\lambda \phi  \tilde \nabla_\lambda \phi ,\eea
where $\tilde \nabla$ are covariant derivatives associated with the 4-dimensional metric ${}_4 \tilde g_{\mu \nu}$. \\
We may rewrite \eqref{iter1a} as 
\bea \label{iter1b}
\int d^4x d^{d-4}y \rgd \ \apd E &=&  \int d^4x d^{d-4}y  \rgdb\ e^{(d-4)\phi} \Big(
{}_4 \tilde E + (d-3)\tilde \nabla_\mu \,{}_4 \tilde J^\mu + (d-3)(d-4) {}_4 \tilde K  \nn && 
+ 2\, {}_4 \tilde R\ \ape R + \ape E-(d-3)\ape R [4\tilde \square \phi + 2(d-4)(\tilde \nabla_\lambda \phi \tilde \nabla^\lambda \phi)] \Big) . \eea
This equation is the starting point in order to proceed with a dimensional reduction of the fields.

\subsubsection{Weyl flat metric in $\mathcal{M}_e$ and DR}\label{example}
In dimensional reduction, as already pointed out, one usually assumes that all the $d$-dimensional fields do not depend on the coordinates of the extra dimensional manifold. The structure of the reduced theory carries symmetries which are decomposed with respect to the original ones. A classical example is that of $\mathcal{N}=1$ supersymmetric Yang-Mills theory in $d=10$, which turns into an $\mathcal{N}=4$ at $d=4$.\\
 In our case, with our metric choice, DR is implemented by assuming that the ${}_e g_{mn}$ metric becomes flat, in order to obtain a pure 4-dimensional integrand. In this case all the external curvatures $_e R$, $_e R_{mn}$ and so on, will obviously vanish.\\
We split the integration measure into $\rgd=\sqrt{{}_4 g} \rge$ and take the flat limit in ${}_e g$ 
($\sqrt{ {}_e{g}}\to 1$). We also reinsert the $\mu^{\epsilon}$ renormalization scale in the original defintion \eqref{ffr}, to obtain

\bea \label{WG in generic case}
\frac{1}{d-4}V_E(g,d) &=&\frac{1}{\epsilon} \left({L}{\mu}\right)^{\epsilon}\int d^4x \rg \  {}_4 \bar{E}+ \left({L}{\mu}\right)^{\epsilon}\int d^4x \rg\ \Big[\phi {}_4\bar{ E}-(4 {} G^{\mu\nu}(\bar\nabla_\mu\phi\bar\nabla_\nu\phi) \nn && +2(\nabla_\lambda \phi \nabla^\lambda \phi )^2 +4\Box\phi \nabla_\lambda \phi \nabla^\lambda \phi ) \Big],
\eea
where all the terms in the integrands are 4-dimensional and $L$ is a space cutoff in the $d-4$ extra dimensions. $L^{\epsilon}$ is the volume of the extra space. Taking the $\epsilon\to 0$ limit, and going back to ordinary 4-$d$ notation, $g_{\mu\nu}=\bar g_{\mu\nu}e^{2 \phi}$, for the fiducial metric, we finally derive the expressions
\beqa
\label{rdef}
\hat{V}'_E( g, \phi)&&=\lim_{d\to 4}\left(\frac{1}{\epsilon}\left(V_E(g,d)-V_E(\bar g,d)\right)\right)\nn
&=&\int d^4x \rg \Big[\phi {}_4 E-(4 {} G^{\mu\nu}(\bar\nabla_\mu\phi\bar\nabla_\nu\phi)+2(\nabla_\lambda \phi \nabla^\lambda \phi )^2 +4\Box\phi \nabla_\lambda \phi \nabla^\lambda \phi ) \Big],
\eeqa
and 
\beq
V'_E=\hat{V}'_E(\bar g, \phi, d) +\log(L \mu)\int d^4x \sqrt{\bar g} \bar E.
\label{more}
\eeq
The dilaton appears with vertices up to order four. The result  reproduces in details the  analysis in 
\eqref{nv2}. In this case we identify the term 
\beq
\frac{\partial V_E(\bar g,d)}{\partial d}\mid_{d=4}=\log(L \mu)\int d^4x \sqrt{\bar g} \bar E
\eeq
which is the Weyl invariant mismatch between the regularization obtained via the WZ action and the one present in $\sm_R$ using standard DR. This example defines a combined DR 
regularization for such types of actions.

A similar analysis can be performed for $V_{C^2}$. In this case, under a Weyl rescaling
\beq
V_{C^2}=\int d^4 x d^{d-4} y \rgd\ \apd C^2=\int d^4 x d^{d-4} y \rgdb\ e^{(d-4)\phi} \apd \bar C^2.
\eeq
As in the previous case, in the $ {}_{(d)}\bar g_{MN}$ metric, the squared curvatures are of the form \eqref{squaredcurvaturefact}, hence the $d$-dimensional Weyl tensor squared is expanded as
\beq \apd \bar C^2 = {}_4 \tilde C^2 + \ape C^2 + \frac{4}{(d-1)(d-2)} {}_4 \tilde R \ape R. \eeq
Then the countertem reads
\beq 
\label{cceq}
\int d^4 x d^{d-4} y \rgd\ \apd C^2=
\int d^4 x d^{d-4} y \rgdb\ e^{(d-4)\phi} \lt {}_4 \tilde C^2 + \ape C^2 + \frac{4}{(d-1)(d-2)} {}_4 \tilde R \ape R \rt . \eeq
Similarly we obtain
\bea 
V_{C^2}=\int d^4 x d^{d-4} y \rgd\ \apd C^2 &=&
\int d^4 x d^{d-4} y \sqrt{ {}_4 \tilde{g}}e^{(d-4)\phi} {}_4\tilde C^2 \nn 
&=& L^{\epsilon} \int d^4 x  \sqrt{ {}_4 \tilde{g}}e^{(d-4)\phi} {}_4\tilde C^2.
\eea
Reinserting the $\mu$ dependence, the expansion in $\epsilon$ of this term generates two contributions
\beq
\frac{1}{d-4} V_{C^2}(d)=\frac{1}{\epsilon}\left(\mu L\right)^\epsilon \int d^4 x  \sqrt{ {g}}{} C^2 +O(\epsilon)
+ \int d^4 x  \sqrt{{g}}\phi C^2,
\eeq
the first of them relevant for the cancellation of the singular behaviour of $\sm_B(d)$ as $d$ goes to 4. Differently from the similar counterterm in $V_E$, this is necessary in order to regulate the divergences of $\sm_B(g,d)$ in the $d\to 4$ limit.   \\
Also in this case, by defining 
\beq
\hat{V}'_{C^2}=\frac{1}{\epsilon}\left( V_{C^2}(g,d)-V_{C^2}(\bar g, d)\right),
\eeq
we have 
\beq
\hat{V}'_{C^2}(\bar g,d)=\int d^d x  \sqrt{{\bar g}}\phi \bar C^2
\eeq
and
\beq
V'_{C^2}(4)=\hat{V}'_{C^2}(\bar g,d) + \log( L \mu) \int d^4 x  \sqrt{{\bar g}}\bar C^2, 
\eeq
giving from \eqref{nv2} 
\beq
\left(2g_{\mu\nu}\frac{\delta \mathcal{S}^{A}}{\delta g_{\mu\nu}}- 2g_{\mu\nu}\frac{\delta \mathcal{S}^{WZ}}{\delta g_{\mu\nu}}\right)= \log( L \mu)\int d^4 x  \sqrt{\bar g}\left(b' \bar E +b\bar C^2\right).
 \eeq
The conformal backreaction, identified in $\sm_R$, can then be expressed in the final form
\beqa
\sm_R &=&\sm_f + b'\int d^4x \rg \Big[\phi {}_4 E-(4 {} G^{\mu\nu}(\bar\nabla_\mu\phi\bar\nabla_\nu\phi)+2(\nabla_\lambda \phi \nabla^\lambda \phi )^2 +4\Box\phi \nabla_\lambda \phi \nabla^\lambda \phi ) \Big]\nn
&& + b\int d^d x  \sqrt{{\bar g}}\phi \bar C^2 + \log( L \mu)\int d^4 x  \sqrt{\bar g}\left(b' \bar E +b\bar C^2\right),
\label{EE}
\eeqa
valid in DR, where the only missing term is $\sm_f$. Both $\sm_f$ and the log-contribution are Weyl-invariant (i.e. $\phi$-independent) terms which are part of the regulated action. If we limit our attention only to a GB theory, with a classical singular rescaling of the GB coupling, as in ordinary $d=2$ gravity \cite{Mann:1992ar},
then $\sm_f$ and the log terms are obviously absent, while at the same time we need to set $b=0$. 
Notice that in \eqref{EE} the logs of the renormalization scale $\mu$ {\em are present} and accompany the $\sqrt{\bar{g}} \bar{C}^2$ density. If, as already discussed, we vary $\bar{g}$ and $\phi$ independently, as we should, then scale invariance is violated.

\section{The quartic dilaton action and the conformal breaking scale $(f)$}
\label{ef}
Extending $\sm_R$ in order to derive a Einsten GB/Weyl theory is quite straightforward, but it is not a unique procedure. At the same time, this theory can be accompanied by other  terms of various types.  One can add, for instance, the EH term. \\
We recall that the EH term may be expressed either in terms of the fiducial metric, as $\sm_{EH}(\bar g,4)$, where 
\beq
\sm_{EH}(\bar g,4)\equiv\int d^4 x \sqrt{\bar g}\left( M_P^2 \bar R + 2\Lambda\right),
\eeq
generating an action at $d=4$ of the form 
\beqa
\sm_{EGBW_1} \equiv \sm_{EH}(\bar g,4) +\sm_R(\bar g,\phi),
\eeqa
with
\beq
\sm_R(\bar g,\phi) = \sm_f(\bar g)  +\sm_A(\bar g,\phi), 
\eeq
and $\sm_A$ given by \eqref{EE},
or, alternatively, by promoting the entire EGBW theory to $d$-dimensions and performing the 
$d\to 4$ limit on all of its components. \\
In this second case, if we perform a Weyl trasformation also on the EH action, we derive the ordinary form of the dilaton gravity action
\bea \int d^d x \rg ( M_P^2 R-2 \Lambda) &=& 
\int d^dx \rgb \  e^{(d-2) \phi} \lt M_P^2 [_d\bar R - 2(d-1)\bar \square \phi - (d-1)(d-2) \bnabla_\lambda \phi \bnabla^\lambda \phi] - 2e^{2\phi} \Lambda \rt \nn &&
=\int d^dx \rgb  e^{(d-2) \phi} \left(M_P^2[\bar R + (d-1)(d-2) \bnabla_\lambda \phi \bnabla^\lambda \phi] - 2e^{2 \phi}\Lambda \rt. \eea
DRed of this action leads to the ordinary dilaton gravity $\sm_{EHd}$ in the Jordan (string) frame
\beq
\label{d1}
\sm_{EHd_1}(\bar g,\phi)=\int d^4x \rgb  e^{2 \phi} \left(M_P^2[\bar R + 6\bnabla_\lambda \phi \bnabla^\lambda \phi] - 2e^{2 \phi}\Lambda \rt. 
\eeq
Logarithmic, scale dependent terms are absent in this action, since we can smoothly take the $d\to 4$ limit in DR from $\sm_{EH}$ in $d$ dimensions, due to finiteness. 
 
We can add $\sm_R$ to this action, obtaining the corresponding EGBW action 
-denoted as  $EGBW_1$ -
\beqa
\sm_{EGBW_1} \equiv \sm_{EH{d_1}}(\bar g,\phi) +\sm_R(\bar g,\phi),
\eeqa
with $\sm_R(\bar g,\phi)$ given by \eqref{EE}.
We have also observed that there are variants of the 
theory in which the logarithmic  $\log(\mu L)$ scale dependent terms are absent from the counterterms. 
We work in the context of this variant, which corresponds to a redefinition of the renormalized quantum effective action $\sm_R$ in the form
\beq
\tilde \sm_R=\sm_f(4) + \sm_{WZ},
\eeq
giving 
\beqa
\label{mod1}
\tilde \sm_{EGBW_1} \equiv \sm_{EH{d_1}}(\bar g,\phi) +\tilde\sm_R(\bar g,\phi).
\eeqa
This theory is defined according \eqref{EE} by the choice $\mu=1/L$, which removes the log terms in DR. \\
We are now going to address one aspect of this dilaton gravity action.\\
One of the most important issues 
concerns  the presence of constrains between the trace equation of motion of the fiducial metric and the equation of the conformal factor, as shown by \eqref{cons}. This relation is induced by the renormalization procedure and is obviously related to the anomaly, that breaks the residual invariance of the conformal decomposition \eqref{mar}. 
Notice that $\phi$ does not carry any dimension and it is clear that a correct normalization of this field requires the introduction of a scale $f$. Therefore, the selection of a given fiducial metric $\bar{g}$ is directly linked to the emergence of $f$, which breaks the conformal symmetry. To investigate this point, we proceed as follows.\\
Before expanding we send $\phi\rightarrow -\phi$, obtaining
\begin{align} \label{EGBW1}
&\tilde S_{EGBW_1} = \frac{1}{16\pi G}\int d^4 x \rg \  e^{-2\phi} \lt[ R + 6\nabla_\lambda \phi \nabla^\lambda \phi ] -2e^{-2\phi}\Lambda\rt  + \sm_f(4) \nn
&+\int d^4 x \rg \biggl[-\phi(b' E+ b C^2)-b'\left(4 G^{\mu\nu}(\nabla_\mu\phi\nabla_\nu\phi)+2(\nabla_\lambda \phi\nabla^\lambda \phi)^2-4\bar\Box\phi\nabla_\lambda \phi\nabla^\lambda \phi\right)\biggl].
\end{align}
We omit, for simplicity, the bar symbol on the gravitational metric.
It is quite straightforward to show that such a Lagrangian describe a spontaneously broken phase, due to the presence of a bilinear mixing between the scalar field $\phi$ and the metric. \\
To show this, it is convenient to introduce the field redefinition 
\beq
e^{-2 \phi}=1- \frac{\tilde{\phi}}{f} \qquad  \phi=-\frac{1}{2} \log(1-\frac{\tphi}{f})
\eeq  
for which the action is rewritten as 

\begin{align} \label{EGBW2}
&\tilde S_{EGBW_1} = \frac{M_P^2}{2}\int d^4 x \rg \left( R - \frac{1}{f}\tphi R + \frac{3}{2}\frac{1}{(1-\frac{\tphi}{f}) f^2}\partial_\lambda \tphi \partial^\lambda \tphi  -2(1-\frac{\tphi}{f})\Lambda \right)  + \sm_f(4) \nn
&+\int d^4 x \rg \Biggl[\frac{1}{2}\log(1-\frac{\tphi}{f})(b' E+ b C^2)- b' \left( G^{\mu\nu}\frac{1}{(1-\frac{\tphi}{f})^2 f^2}(\partial_\mu\tphi\partial_\nu\tphi)+\frac{1}{8(1-\frac{\tphi}{f})^4 f^4}(\partial_\lambda\tphi \partial^\lambda\tphi)^2\right.\nn
&\qquad \qquad \left. -\frac{1}{2(1-\frac{\tphi}{f})^3 f^3}\Box_0\tphi\partial_\lambda\tphi\partial^\lambda \tphi -\frac{1}{2(1-\frac{\tphi}{f})^4 f^4}\partial_\mu\tphi\partial^\mu\tphi \partial_\nu\tphi\partial^\nu\tphi +
 \frac{1}{2(1-\frac{\tphi}{f})^2 f^3} \Gamma^\lambda\partial_\lambda \tphi \partial_\sigma\tphi\partial^\sigma \tphi \right)\Bigg],
\end{align}

where $M_P^2=1/(8 \pi G_N)$ is the reduced Planck mass and $\Gamma^\lambda\equiv g^{\mu\nu}\Gamma^\lambda_{\mu\nu}\equiv-\frac{1}{\sqrt{g}}\partial_\mu(\sqrt{g}g^{\lambda \mu}) $. 
Notice that the coupling of the action can be organized in terms of 
interactions of increasing mass-dimensions in an expansion in $\tphi/f$. The presence of a bilinear mixing 
in the EH part of the effective action $(\sim M_P^2/f )\tphi R$ is indicating that we are describing a broken phase.  A solution of the equations of motion can be obtained by setting 
 $\phi$ constant, and taking a flat fiducial metric $\bar{g}_{\mu\nu}=\delta_{\mu\nu}$, (i.e. a Weyl flat $g_{\mu\nu}$). In this case 
\beq
\phi=v, \qquad R_{\mu\nu}=\frac{1}{4}R g_{\mu\nu}, \qquad R=24 \lambda v^2.
\eeq
An alternative approach is to proceed by introducing a different field redefinition of the form 
\beq
e^{-\phi}=\bar{\chi}(x),
\eeq
where $\bar{\chi}(x)$ can be related to a mass dimension-1 scalar as $\bar{\chi}(x)={\chi(x)}/({\sqrt{3}f})$
generating the coupling 
\beq
\mathcal{L} \supset \frac{M_P^2}{2 f^2}\sqrt{g}\left( \frac{1}{2}g^{\mu\nu}\partial_{\mu}\chi\partial_{\nu}\chi +\frac{1}{6}R \chi^2\ldots\right). 
\eeq
It is easy to realize that the $R\chi^2$ term carries the wrong sign, since for slowly varying curvature behaves essentially as a mass term 
with $m^2\sim R$. The presence either of mixing terms or of mass terms with the wrong sign are the signatures that the procedure of Weyl gauging generates a Lagrangian in a broken phase. \\
Concerning the asymptotic structure of $\tilde\sm_{EGBW_1}$, it is convenient to organize the terms appearing in it as an expansion in the two scales $1/f$ and $1/(f^n M_P^2)$, obtaining 
\beqa
\tilde\sm_{EGBW_1}&=&\frac{M_P^2}{2}\int d^4 x \sqrt{g}\left( R -\frac{\tphi}{\bar{f}} R +
\frac{1}{2\bar{f}^2}(\partial_\mu\tphi)^2   + 2\frac{\tphi}{\bar{f}}\Lambda -2 \Lambda  + O(1/\bar{f}^2) \right.\nn
&&\left. \qquad \qquad \qquad -\frac{\tphi}{\bar{f}M_P^2}(\alpha E + \alpha' C^2) +O(1/(\bar{f}^2 M_P^2) \right),
\eeqa

where we have redefined $\tphi\to \tphi/\sqrt{3}$ and $\bar{f}=\sqrt{3} f$.

At large $\bar{f}$, with $f \ll M_P$,  the dilaton field can be expressed in terms of the fiducial metric using the nonlocal relation 
\beq
\label{limit}
\tphi\sim \frac{1}{\Box}\left( - \bar{f}({R} + {\Lambda}) -\frac{\bar{f}}{M_P^2}(b'  E +b  C^2)\right),
\eeq
where the asymptotic expression of the field can be removed on-shell, via an auxiliary nonlocal interaction 
containing suppressed - by $\bar{f}/ M_P^2$ -  nonlocal couplings to the anomaly ($\frac{1}{\Box}(\alpha E +\alpha' C^2)$) and to the curvature ($\frac{1}{\Box}R$). If the conformal breaking scale $f$ grows 
towards $M_P$, the leading behaviour of the dilaton contribution is described by $\frac{1}{\Box}R$.
We have seen that the structure of the effective action depends on the way we perform the $d\to 4$ limit and on the choice of the metric that is used for evaluating the finite contributions $V'_E$ and $V'_{C^2}$ or $\hat V'_E$ and $\hat V'_{C^2}$, resulting from the renormalization procedure.\\ 
Eq. \eqref{limit} indicates that after the breaking of the conformal symmetry, here simply introduced by a renormalization procedure that accounts for the anomaly, the coupling of the dilaton to the anomaly $\mathcal{A}(x)$ is suppressed, compared to its coupling to the curvature or to the cosmological constant. 
One may observe that the enhanced coupling of such field to $R$, is a result of the Weyl gauging of the EH action, and it is not related to the inclusion of the quantum corrections. It shows up as a purely classical effect, which is expected to be present in any dilaton gravity model, given the generality of the procedure.

\section{Moving towards the UV: The reconstruction at $d=4$  for GB }\label{UV}
The reconstruction of the anomaly action in $d=4$ follows the standard procedure introduced long ago by Riegert \cite{1984PhLB..134...56R}, that we will review and extend to the GB case, in order to underline the difference between the various possible effective actions that may follow. 
Therefore, the regularization of the GB term can indeed generate regulated GB actions which can either take a local or a nonlocal form, depending on the way the conformal factor is treated in the regularization procedure \cite{Mazur:2001aa,Coriano:2022knl}.\\
 Both $V'_E$ and $V'_{C^2}$ are {\em local} expressions of the fiducial metric 
$\bar{g}_{\mu\nu}$ and of the field $\phi$. As already pointed out, their nonlocal structure will be apparent only if we are able to remove $\phi$, by re-expressing it in terms of the original metric $g_{\mu\nu}$, and this is not always possible. The case discussed by Riegert is one in which the conformal scaling relation \eqref{form1} is linear in $\phi$, and the dilaton can be removed by an integration procedure. This is a consequence of the fact that the rescaling is performed at $d=4$. 
Indeed, in this case the rescaling gives
\beq
\sqrt{g}\,\Big(E-\frac23\,{\Box} R\Big)\,=\,
\sqrt{\bar g}\,\Big({\bar E}-\frac23\,{\bar \Box} {\bar R}
+ 4{\bar \Delta_4}\phi \Big)\,,
\label{119}
\eeq
where $\Delta_4$ is the fourth order self-adjoint 
operator, which is conformal invariant when it acts on a scalar function of vanishing scaling dimensions 
\beq
\Delta_4 = \na^2 + 2\,R^{\mu\nu}\na_\mu\na_\nu - \frac23\,R{\Box}
+ \frac13\,(\na^\mu R)\na_\mu\,.
\label{120}
\eeq
and satisfies the relation
\beq
\sqrt{-g}\,\D_4\chi_0=\sqrt{-\bar g}\,\bar{\D}_4 \chi_0,\label{point2}
\eeq
if $\chi_0$ is invariant (i.e. has scaling equal to zero) under a Weyl transformation. \\
Eq. \eqref{119} is crucial for the elimination of $\phi$ from the effective action. This is obtained by the 
inclusion of a boundary term $(\Box R)$.  It is clear that the identification of the anomaly action using this equation does not follow the approaches outlined in the previous sections, which are entirely based on DR and the choice of appropriate metrics and manifold of integrations. 
Riegert's approach can be modified by turning to $d$ dimensions, with the inclusion of finite renormalizations and rendered consistent with DR. 
There are variants of $E$ that can be introduced in order to satisfy \eqref{ep2} and allow to 
eliminate $\phi$, quite closely to \eqref{119}, as we are going to discuss in the sections below and our goal will be to 
propose the same procedure also for 4EGB theories.  
We can introduce for instance the modified and extended version of $E$ in the form 

\beq
\label{ext}
E^{ext}\equiv E +\frac{\epsilon}{2(d-1)^2}R^2,
\eeq
which is useful in order to investigate the contribution of the $V_E$ counterterm - and of its variants -  to the effective action. Notice that the two extra terms that appear on the rhs of \eqref{ext}, correspond to a boundary contribution $(\Box R)$, and to an $O(\epsilon)$ modification $(\sim \epsilon R^2)$ that vanish if we ensure either trivial boundary conditions on the metric, or we perform the $d\to 4$ limit. $E^{ext}$ plays a role in the identification of a form of the effective action which is quite close to Riegert's action. \\
The scaling relation \eqref{119} is rather unusual, in the sense that its metric variation links boundary terms in the two metrics $g_{\mu\nu}$ and $\bar{g}_{\mu\nu}$. One can show that in $d=4$, under a metric variation $\delta$

\beq
\label{intt}
\frac{1}{4}\delta (\sqrt{g} E)=\sqrt{g}\,\nabla_\sigma \delta X^\sigma,
\qquad 
\delta X^\sigma=\varepsilon^{\mu\nu\alpha \beta}\varepsilon^{\sigma\lambda\gamma\tau}
\delta \Gamma^\eta_{\nu\lambda}g_{\mu\eta}R_{\alpha \beta \gamma \tau},\qquad \varepsilon^{\mu\nu\alpha \beta}=\frac{\epsilon^{\mu\nu\alpha \beta}}{\sqrt{g}}
\eeq

\beq\delta( \sqrt{g}\Box R)=\sqrt{g}\Box \delta\zeta,\qquad  \delta\zeta=-R^{\mu\nu}\delta g_{\mu\nu} +\nabla^\mu\nabla^\nu \delta g_{\mu\nu} 
-\Box(g^{\mu\nu}\delta g_{\mu\nu}).
\eeq

These relations follow after some integration by parts, having observed that the conformal factor varies like a scalar under the Weyl rescalings in two different frames $x$ and $x'$. This results from the fact that a fiducial metric transforms as an ordinary tensor in the two frames, hence 
\beq
g_{\mu\nu}(x)=\bar{g}_{\mu\nu}(x) e^{2 \phi(x)}  \qquad g'_{\mu\nu}(x')=\bar{g}'_{\mu\nu}(x') e^{2 \phi'(x')}
\qquad \phi'(x')=\phi(x).
 \eeq 
It is convenient to define 
\beq
\delta \Sigma^\sigma=\sqrt{g}g^{\sigma\beta}\partial_\beta \delta\zeta \qquad
\label{int1}
\eeq and vary 
both sides of \eqref{119} to obtain, using 
\beq
\delta_\phi \left(\sqrt{g}\Box R\right) = \epsilon \delta\phi \Box R +(d-6)\sqrt{g}\nabla^\lambda R \nabla_\lambda \delta\phi -2 \sqrt{g}R\nabla^2\delta \phi -2 (d-1)\sqrt{g}\nabla^4\delta \phi
\eeq 
the scaling relation at $d=4$
\begin{align}
& \delta_\phi\left(\frac{1}{4}\sqrt{-g}\left(E-\frac{2}{3}\square\, R\right)\right)=\sqrt{- g}\D_4\delta\phi, 
\label{pointd}
\end{align}

which simplifies in the form 

\beq
\partial_\sigma\left( \delta_\phi X^{\sigma} -\frac{1}{6}\delta_\phi\Sigma^\sigma\right)=\sqrt{- g}\D_4\delta\phi, 
\label{cc}
\eeq
if we use \eqref{intt} to relate it to a boundary contribution. Here we have used the general variation 
of $\delta X^\sigma$ specialised to changes in the dilaton field ($\delta_\phi$)
\beq
\delta_\phi X^\sigma=\varepsilon^{\mu\nu\alpha \beta}\varepsilon^{\sigma\lambda\gamma\tau}
\delta_\phi \Gamma^\eta_{\nu\lambda}g_{\mu\eta}R_{\alpha \beta \gamma \tau}.
\eeq
We have defined
\beq
\delta_\phi\Gamma_{\mu\nu}^\lambda=\de_\mu^\lambda \nabla_\nu\delta\phi+\de^\lambda_\nu \nabla_\mu\delta\phi- e^{-2 \phi} g_{\mu\nu}\nabla^\lambda\delta\phi,
\eeq
derived from \eqref{dg}, using $\bar\nabla_\mu\delta \phi=\nabla_\mu\delta\phi$ on scalars,
while and analogous variation $\delta\Sigma^\sigma$ in \eqref{int1} is specialised in the form 
\beq
\delta_\phi \Sigma^\sigma=\sqrt{g}g^{\sigma\beta}\partial_\beta \delta_\phi\zeta,  
\eeq
where
\beq
  \delta_\phi\zeta=-R^{\mu\nu}\delta_\phi g_{\mu\nu} +\nabla^\mu\nabla^\nu \delta_\phi g_{\mu\nu} 
-\Box(g^{\mu\nu}\delta_\phi g_{\mu\nu}),   \qquad  \textrm{with}\qquad   \delta_\phi g_{\mu\nu}=2g_{\mu\nu}\delta\phi.
\eeq
Eq. \eqref{cc}, integrated over spacetime, gives consistently 
\beq
\int d^4 x \partial_\sigma\left( \delta_\phi X^{\sigma} -\frac{1}{6}\delta_\phi\Sigma^\sigma\right)=0,
\eeq
if we assume asymptotic flatness, and therefore
\beq
\int d^4 x \sqrt{- g}\D_4\delta\phi=0,
\eeq
that follows from the self-adjointness of $\Delta_4$
\begin{equation}
\int d^4x\sqrt{-g}\,\y(\D_4\x)=\int d^4x\sqrt{-g}\,(\D_4\y)\x\label{point3},
\end{equation}
where $\x$ and $\psi$ are scalar fields of zero scaling dimensions. 

The scaling relation $\eqref{119}$ is strictly valid at $d=4$ and clearly is much simplified compared to 
\eqref{form1}, which is valid in $d$ dimensions.
Clearly, Eq. \eqref{119} is not directly related to a DR procedure, but simply takes the expression of the anomaly as a given fundamental 4-dimensional result and integrates out the 
dilaton field from the scaling relation, to derive the nonlocal form of the action.  \\
We are now going to briefly review this point. \\
It is convenient to redefine \eqref{119} in the form 
\beq
 J(x)=\bar{J}(x) + 4 \sqrt{g}\Delta_4\phi(x),\qquad     \bar J(x)\equiv \sqrt{\bar g}\left( \bar E-\frac{2}{3}\bar \Box \bar R\right), \qquad  J(x)\equiv \sqrt{ g}\left(  E-\frac{2}{3} \Box  R\right) 
 \eeq
\begin{equation}
(\sqrt{-g}\,\D_4)_xD_4(x,y)=\d^4(x,y).\label{point4}
\end{equation}

We invert \eqref{119} using the properties of the operator $\Delta_4$ to find the explicit form of the function $\phi(x)$, obtaining 
\begin{equation}
\label{onshell}
\phi(x)=\frac{1}{4}\int d^4y\,D_4(x,y)(J(y)- \bar{J}(y)).
\end{equation}
This sets $\phi$ on-shell. The derivation of $\sm_{WZ}$ requires the solution of the equation 
\beq
\frac{\delta \mathcal{S}_{WZ}^{(GB)}}{\delta \phi}=J,
\eeq
clearly identified in the form
\beq
\sm_{WZ}=\int d^4 x \sqrt{\bar g}\left(\bar J \phi + 2 \phi \Delta_4 \phi\right).
\eeq
At this stage it is just matter of inserting the on-shell expression of $\phi$ \eqref{onshell} into this equation to obtain 
the WZ action, in the form
\beq
\sm_{WZ}=\sm_{anom}(g)- \sm_{anom}(\bar g),
\label{WW}
\eeq
with 
\beq
\sm_{anom}(g)=\frac{1}{8}\int d^4 x d^4 y J(x) D_4(x,y) J(y),
\eeq
and a similar expression for $\sm_{anom}(\bar g)$. 
Using the explicit expression of $\phi$, and including the contributon from the rescaled $C^2$ term, we finally find the nonlocal and covariant anomaly effective action as
\begin{equation}
\mathcal{S}_{\rm anom}^{}(g) =\frac {1}{8}\!\int \!d^4x\sqrt{-g_x}\, \left(E - \frac{2}{3}\sq R\right)_{\!x} 
\int\! d^4x'\sqrt{-g_{x'}}\,D_4(x,x')\left[\frac{b'}{2}\, \left(E - \frac{2}{3}\sq R\right) +  b\,C^2\right]_{x'}.
\label{Snonl}
\end{equation}

\section{Modified Euler density and the nonlocal GB action} 
\label{mod}
Notice that if the rescaling is performed at $d=4$, and the extra field $\phi$ is reabsorbed into the definition 
of $g_{\mu\nu}$, giving a nonlocal action, then no scale of expansion is present in  $\sm_A(4)$. If we move away from 4 dimensions, and this is clearly allowed in DR, then it is obvious that extra components of the metric will be present in the expressions of $V'_E$ and $V'_{C^2}$, and the computation of the effective action will be affected by the choice of the fiducial metric over which we integrate in $d$ dimensions. \\
In the context of DR and, in particular, in the analysis of the effective actions, it is clear that variants of the topological terms are possible.\\
The functional differential form $V_E$, constructed out of $E_4\equiv E$ is not the only possible one. It is clearly exact since 
\beq
\label{abv}
\delta_\phi (\sqrt{g} E_4)=\epsilon \sqrt{g}E_4 \delta\phi,
\eeq
that can be verified directly by taking the trace of $V_E^{\mu\nu}$. Another way is to introduce separately the variations 
\beq
\delta_{\phi} \left(\sqrt{g}R^2\right)=\delta\phi \left(\epsilon\sqrt{g}R^2 -4 (d-1)\sqrt{g}\Box R\right),
\eeq
giving under integration 
\beq
\delta_\phi\int d^d x \sqrt{g}R^2=\epsilon\sqrt{g}R^2 -4 (d-1)\sqrt{g}\Box R.
\eeq
Similarly
\beq
\delta_\phi \left(\sqrt{g} (R_{\mu\nu\alpha\beta})^2\right) =\delta\phi\left(\epsilon \sqrt{g}  (R_{\mu\nu\alpha\beta})^2 
-8 \sqrt{g}\nabla_\mu\nabla_\nu R^{\mu\nu}
\right),
\eeq

\beq
\delta_\phi \left(\sqrt{g} (R_{\mu\nu})^2\right)=\epsilon \delta\phi \sqrt{g} (R_{\mu\nu})^2 -2 \sqrt{g}\Box R \delta\phi -2(d-2) \sqrt{g} \nabla_\mu\nabla_\nu R^{\mu\nu}\delta \phi,
\eeq
Obviously, as we move away from $d=4$, modifications of such densities are possible.\\
 In general, we can modify such forms either by boundary terms, which play a role only if we include a spacetime boundary and/or by additional diffeomorphism invariant contributions of $O(\epsilon)$. \\
If we consider the extended expression of $E_4$ given by \eqref{ext},
in this case we define the counterterm
\beq
\tilde{V}_{E}=\int d^d x\sqrt{g} \left(E_4 + \epsilon\frac{R^2}{2 (d-1)^2}\right).
\eeq

Using the variations above, one obtains

\beq
\delta_\phi (\sqrt{g} E_{ext})=\delta\phi\epsilon\left(\sqrt{g}E_{ext}-\frac{2}{d-1}\sqrt{g}\Box R\right),
\eeq 
giving under integration 

\beq
\delta_\phi\int d^d x \sqrt{g} E_{ext}=\epsilon \sqrt{g}(E_{ext}- \frac{2}{d-1} \Box R). 
\eeq
Also in this case one needs to be careful about the $d\to 4$ limit since the metric is still $d$-dimensional 
and one has to proceed with an accurate definition of the corresponding invariants. One possibility is to perform a dimensional reduction as already discussed.  This introduces a cutofl 
$L^\epsilon$ that can be consistently removed as $\epsilon\to 0$. 

\subsection{The nonlocal EGB expansion} 
One of the standing issues concerning the consistency of the scaling approach introduced in \eqref{119} is that it is possible to make it consistent with DR, promoting to $d$ dimensions from $d=4$. This point can be addressed and solved by a redefinition of $S_{WZ}$
using $E_{ext}$ with $V_E\to \tilde{V}_E$, obtaining 
\beq
\tilde{V}_E=\int d^d x \sqrt{g}E_{ext}\, , 
 \eeq

\begin{equation}
\mathcal{S}^{(WZ)}_{GB} =\frac{\alpha}{\epsilon}\left(\tilde{V}_E(\bar{g}_{\mu\nu}e^{2\phi},d)- \tilde{V}_E(\bar{g}_{\mu\nu},d\right).
\label{inter}
\eeq

To derive its nonlocal expression, we can use the relation
\beq
\frac{\delta}{\delta\phi}\frac{1}{\epsilon}\tilde{V}_E(g_{\mu\nu},d)= \sqrt{g}\left(E-\frac{2}{3}\Box R +
\epsilon\frac{R^2}{2(d-1)^2}\right)
\eeq
in \eqref{inter}, to obtain  
 \beqa
 \frac{\delta \mathcal{S}^{(WZ)}_{GB}}{\delta\phi}&=&\alpha\sqrt{g}\left(E-\frac{2}{3}\Box R \right)\nonumber \\
&=&\alpha\sqrt{\bar g}\left(\bar E-\frac{2}{3}\bar \Box\bar R + 4 \bar\Delta_4 \phi\right),
\label{solve}
\eeqa
and henceforth
\begin{equation}
\mathcal{S}^{(WZ)}_{GB} = \alpha\int\,d^4x\,\sqrt{-\bar g}\,\left\{\left(\overline E - {2\over 3}
\bar{\Box} \overline R\right)\phi + 2\,\phi\bar\Delta_4\phi\right\},\,
\label{WZ2}
\end{equation}

As before, we can solve for $\phi$, deriving the regulated GB action
\beqa
 \mathcal{S}^{(WZ)}_{GB}& =& 
{\alpha\over 8} \int d^4x\,\sqrt{-g}\, \int d^4x'\,\sqrt{-g'}\,
\left(E_4 - {2\over 3} \Box R\right)_x\, \nonumber \\
&&\qquad \times D_4(x,x')\left(E- {2\over 3} \Box R\right)_{x'},\,
\label{anomact}
\eeqa
that coincides with the result provided in \cite{Mazur:2001aa} by Mazur and Mottola. 

The nonlocal EGB action can be expanded, at least around a flat spacetime, in terms of the combination of the product of scalar curvature $R$ and the inverse of the D'Alembertian of flat space, i.e. of the variable $R\Box^{-1}$, which is dimensionless \cite{Coriano:2018bsy,Coriano:2017mux,Coriano:2021nvn}. This results both from perturbative computations performed around flat space and from studies of the hierarchical structure of the CWIs. \\
At this stage, we are ready extract the classical interactions present in the action by an expansion around flat space. \\
One rewrites the nonlocal anomaly action in an equivalent local form  
\bea
\label{loc}
&&\hspace{-1.5cm} \cS_{\rm anom}(g,\vf) \equiv -\sdfrac{1}{2} \int d^4x\,\sqrt{-g}\, \Big[ (\sq \vf)^2 - 2 \big(R^{\m\n} - \tfrac{1}{3} R g^{\m\n}\big)
(\na_\m\vf)(\na_\n \vf)\Big]\nn
&& \hspace{1.5cm} +\, \sdfrac{1}{2}\,\int d^4x\,\sqrt{-g}\  \Big[\big(E - \tfrac{2}{3}\sq R\big)  \Big]\,\vf,
\label{Sanom}
\eea
that can be varied with respect to $\phi$, giving
\be
\sqrt{-g}\,\D_4\, \vf = \sqrt{-g}\left[\sdfrac{E}{2}- \sdfrac{\!\sq R\!}{3} \right] \label{phieom}.
\ee
The metric can be expanded perturbatively in the form 
\bes
\bea
g_{\m\n} &=& g_{\m\n}^{(0)} + g_{\m\n}^{(1)} + g_{\m\n}^{(2)} + \dots \equiv \eta_{\m\n} + h_{\m\n} + h_{\m\n}^{(2)} + \dots\\
\vf &=& \vf^{(0)} +  \vf^{(1)} +  \vf^{(2)}  + \dots
\eea
\ees
The expansion above should be interpreted as a collection of terms generated by setting 
\beq
g_{\m\n} = \delta_{\mu\nu} + \kappa h_{\m\n} 
\eeq
having reinstated the coupling expansion $\kappa$, with $h$ of mass-dimension one,  
and collecting all the higher order terms in the functional expansion of \eqref{loc} of the order $h^2$, $h^3$ and so on. A similar expansion holds for $\vf$ if we redefine $ \vf^{(1)}=\kappa \bar\vf^{(1)},  \vf^{(2)}=\kappa^2 \bar \vf^{(2)}$ and so on. One obtains the relations
\bes
\bea
&&\hspace{5cm}\sqb^2 \vf^{(0)} = 0 \label{eom0}\\
&&\hspace{-1.5cm}(\sqrt{-g} \D_4)^{(1)} \vf^{(0)} + \sqb^2 \vf^{(1)} = \left[\sqrt{-g}
\left( \sdfrac{E}{2}- \sdfrac{\!\sq R\!}{3} \right)\right]^{(1)}
= - \sdfrac{\!1\!}{3}\, \sqb R^{(1)} \label{eom1}\\
&&\hspace{-2cm}(\sqrt{-g} \D_4)^{(2)} \vf^{(0)} + (\sqrt{-g} \D_4)^{(1)} \vf^{(1)} + \sqb^2 \vf^{(2)} =
\left[\sqrt{-g}\left(\sdfrac{E}{2}- \sdfrac{\!\sq R\!}{3}  \right)\right]^{(2)} \nn
&&\hspace{5.5cm}= \sdfrac{1}{2}E^{(2)} - \sdfrac{1}{3}\, [\sqrt{-g}\sq R]^{(2)},  \label{eom2}
\eea
\ees
where $\sqb$ is the d'Alembert wave operator in flat Minkowski spacetime, and we have used the fact that $E$ and $C^2$ are second order in the fluctuations while the Ricci scalar $R$ starts at first order

\be
\vf^{(1)} = - \sdfrac{\!1 \!}{3\sqb}\, R^{(1)}
\label{vf1}
\ee
and the solution of (\ref{eom2}) is
\be
\vf^{(2)} = \sdfrac{1}{\sqb^2} \left\{ (\sqrt{-g} \D_4)^{(1)}\sdfrac{\!1 \!}{3\sqb} \, R^{(1)} 
+  \sdfrac{1}{2}E^{(2)} - \sdfrac{1}{3}\, [\sqrt{-g}\sq R]^{(2)} \right\}.
\label{vf2}
\ee
In this way we obtain the quadratic term
\be
\cS_{\rm anom}^{(2)} = - \sdfrac{1}{2} \,\int d^4x \, \vf^{(1)} \sqb^2 \vf^{(1)} + \sdfrac{1}{2}\,\int d^4x \, \left( - \sdfrac{2}{3} \sqb R^{(1)}\right) \vf^{(1)}
= \sdfrac{1}{18} \,\int d^4x \, \left(R^{(1)}\right)^2,
\label{Sanom2}
\ee
which is purely local, since all propagators cancel. 
The third order terms in the expansion of the anomaly action are
\bea
\cS_{\rm anom}^{(3)} &=&  - \sdfrac{1}{2} \int d^4x \, \left\{2\,\vf^{(1)} \sqb^2 \vf^{(2)} +\vf^{(1)} \big(\sqrt{-g} \D_4\big)^{(1)} \,\vf^{\!(1)} \right\}\nn
&&\hspace{-1cm} + \sdfrac{1}{2} \int d^4x \left\{\left( - \sdfrac{2}{3} \sqb R^{(1)}\right) \vf^{(2)} + \left(E^{(2)} - \sdfrac{2}{3}\, \sqrt{-g}\sq R\right)^{\!(2)} \vf^{(1)} \right\}.
\label{Sanom3a}
 \eea
The remaining terms in (\ref{Sanom3a}) yield
\bea
&&\cS_{\rm anom}^{(3)} =- \sdfrac{1}{18} \int d^4x \, \left\{R^{(1)}\sdfrac{1}{\sqb} \big(\sqrt{-g} \D_4\big)^{\!(1)} \,\sdfrac{1}{\sqb} R^{(1)} \right\}
- \sdfrac{b'}{6} \int d^4x \left(E- \sdfrac{2}{3}\, \sqrt{-g}\sq R\right)^{\!(2)}\,\sdfrac{1}{\sqb} R^{(1)}.\nn
&&\hspace{2cm}. 
\eea
In the variation of $\Delta_4$ it is convenient first to rewrite 
\beq
\Delta_4= \Box^2 +2 \nabla_\mu(R^{\mu\nu}\nabla^\nu) -\frac{2}{3}\nabla_\mu(R \nabla^\mu),
\eeq
having used the Leibnitz rule and the derivative Bianchi identity $\nabla_\mu R^{\mu\nu}=1/2 \nabla^{\nu} R$. An expansion of this operator to first order in $\delta g_{\mu\nu}$ gives

\be
\big(\sqrt{-g} \D_4\big)^{\!(1)} =  \big(\sqrt{-g} \sq^2\big)^{\!(1)} + 2\, \pa_{\m} \left(R^{\m\n} - \sdfrac{1}{3} \eta^{\m\n} R\right)^{\!(1)}\pa_{\n}.
\ee
An integration by parts gives 
\bea
&&\hspace{-1.1cm}\cS_{\rm anom}^{(3)}\! =\!- \sdfrac{1}{18}\! \int\! d^4x \left\{\!R^{(1)}\!\sdfrac{1}{\sqb} \big(\sqrt{-g} \sq^2\big)^{\!(1)}\sdfrac{1}{\sqb} R^{(1)}\! \right\}
+ \sdfrac{1}{9}\! \int\! d^4x \left\{\!\pa_{\m} R^{(1)}\!\sdfrac{1}{\sqb} \! \left(\!R^{(1)\m\n}\! - \!\sdfrac{1}{3} \eta^{\m\n} R^{(1)}\!\right)\!
\sdfrac{1}{\sqb}\pa_{\n} R^{(1)}\!\right\}\hspace{-5mm}\nn 
&&\hspace{-8mm} - \sdfrac{1}{6}\! \int\! d^4x  E^{\!(2)} \sdfrac{1}{\sqb}R^{(1)}
+ \sdfrac{1}{9} \!\int\! d^4x\,  R^{(1)}  \sdfrac{1}{\sqb} \left(\sqrt{-g}\sq\right)^{\!(1)}R^{(1)}
+ \sdfrac{1}{9} \! \int\! d^4x\, R^{\!(2)}R^{(1)},
\label{Sanom3b}
\eea
which contains only single propagator poles. 
At this stage, using the covariant equation $\nabla_\mu\sqrt{g}=0$ on the tensor density $\sqrt{g}$ we rewrite 
\beq
\left(\sqrt{g}\Box^2\right)^{(1)}\equiv \delta\left(\sqrt{g}\Box^2\right)=\left(\frac{(\sqrt{g}\Box)^2}{\sqrt{g}}\right)^{(1)}\qquad \delta\left(\frac{1}{\sqrt{g}}
\right)(\sqrt{g}\Box)^2=-\delta(\sqrt{g})\Box^2,
\eeq
and
\beq
\left(\sqrt{g}\Box^2\right)^{(1)}=-\delta(\sqrt{g})\Box^2 +\frac{1}{\sqrt{g}}\delta(\sqrt{g}\Box)(\sqrt{g}\Box)
+\frac{1}{\sqrt{g}}(\sqrt{g}\Box)\left(\delta(\sqrt{g}\Box)\right),
\eeq
that in the flat limit becomes
\beq
\left(\sqrt{g}\Box^2\right)^{(1)}=-(\sqrt{g})^{(1)}\sqb^2 + (\sqrt{g}\Box)^{(1)}\sqb + \sqb(\sqrt{g}\Box)^{(1)},
\eeq 
obtaining finally
\bea
&&\hspace{-5mm} \cS_{\rm anom}^{(3)} =
 \sdfrac{1}{9} \int\! d^4x \int\!d^4x'\!\int\!d^4x''\!\left\{\big(\pa_{\m} R^{(1)})_x\left(\sdfrac{1}{\sqb}\right)_{\!xx'}  
 \!\left(R^{(1)\m\n}\! - \!\sdfrac{1}{3} \eta^{\m\n} R^{(1)}\right)_{x'}\!
\left(\sdfrac{1}{\sqb}\right)_{\!x'x''}\!\big(\pa_{\n} R^{(1)})_{x''}\right\}\nn
&&\hspace{-6mm}- \sdfrac{1}{6}\! \int\! d^4x\! \int\!d^4x'\! \left(\, E^{\!(2)}\right)_{\!x}\! \left(\sdfrac{1}{\sqb}\right)_{\!xx'} \!R^{(1)}_{x'}
 + \sdfrac{1}{18} \! \int\! d^4x\, R^{(1)}\left(2\, R^{\!(2)} + (\sqrt{-g})^{(1)} R^{(1)}\right),
\label{S3anom3}
\eea
where the last term is purely local. This action describes graviton interactions up to trilinear fluctuations in the graviton field. \eqref{Sanom} can be expanded, with some extra effort, to quartic and higher orders, providing a definition of the EGB theory, now in a completely nonlocal form and without a dilaton. The appearance of the Green's function of a $\Box^{-1}$ operator, once the expansion is performed around flat space, has been shown in \cite{Giannotti:2008cv,Armillis:2009pq,Coriano:2018zdo} for 3-point functions in the $TJJ$ case and in \cite{Coriano:2017mux} for the 3-graviton vertex TTT. Investigations of this action up to 4-point vertices is underway and results will be presented elsewhere.\\

\section{Conclusions}
In this work we have presented a discussion of the 
structure of the effective action and of its renormalization in DR in some detail, illustrating the main features of the procedure  that allow to identify its explicit expression. We have shown that the regularization, in general, depends on the way we select the background metric, and the integration of the counterterms induces new scales in the gravitational effective action. We have selected to ways to perform the subtractions, one of them giving scale invariant and the other scale non-invariant effective actions.\\
One of the objectives of this analysis has been to establish a link between the anomalous actions in which the dilaton is manifest and those in which the dilaton is removed.
The latter, originally derived by Riegert by a rescaling at $d=4$ of the metric, can be reconsidered in a complete DR scheme, by a redefinition of the topological density, as shown in the crucial analysis of Mazur and Mottola. Our analysis differs from previous ones since it focuses on the entire action, and not only on its anomaly related parts, or just the WZ part of the same action, retracing step by step the renormalization procedure. 

As discussed in the previous sections, Weyl invariant terms are essential, since they are responsible for the breaking of scale invariance and are, as such, part of the anomaly, though not of the trace anomaly.\\
 From this point of view, the reduction of the anomaly induced action only to the Weyl-variant contributions, does not offer a complete view over the breaking of scale invariance. As we have shown, Weyl invariance should carefully restated as an invariance under a joint variation of 
the fiducial metric $and$ of the dilaton field, that we have called "$\sigma$ variation", beside the usual requirement of invariance under a rescaling of the entire metric. Such variation is clearly associated with a symmetry which resembles Stuckelberg's trick for a massive spin-1 (Abelian) gauge theory, in this case broken by the renormalization procedure.\\
 This breaking should be interpreted as the signature of violation of dilatation invariance by a certain part of the action. It may not be present in certain schemes, al least in some parts of the complete action, as in the WZ case, but they appear in others.\\
Therefore, a specific part of the action may be $\phi$-independent, and may not contribute to the trace anomaly, but it can still carry logs of a renormalization scale. \\
As with any regularization, we have the right to carry out a finite renormalization of the effective action, by means of a finite modification of the counterterms. In this case, they modify the scheme dependent part of the trace anomaly equation, and may allow the dilaton to be removed from the spectrum. For this reason, the action may take either a local or non-local form.
We have argued that nonlocal actions are suitable for a description of the physical implications of the conformal anomaly in the UV, and may play an important role in the analysis of the conformal backreaction close to the Planck scale. On the countrary, local actions are suitable for describing the effect of the anomaly at a certain (lower) scale, $f$. \\
We have also shown that a $R\Box^{-1}$ behaviour emerges from the quartic (local) dilaton action as we push the breaking scale up towards the Planck scale, once we couple the theory to the EH action. The 
analogous behaviour identified in the nonlocal action, where the same parameter $R\Box^{-1}$ characterizes the expansion around flat space, is however due only to the anomaly part of this action and, differently from the quartic dilaton gravity case, it is not related to the inclusion of the EH term. Indeed the two actions are different even if we neglect contributions of $O(1/f^3)$ in the expression of the local quartic action, limiting the theory to a quadratic dilaton. \\  
Finally, we have also shown how our results on the structure of $\sm_R$ impacts a class of theories, the EGB theories, in which a classical singular limit on the Gauss-Bonnet term is performed. These quartic dilaton gravity theories have recently received significant attention, for defining local actions of Horndeski type.\\
 We have shown that the relation between the equation of motion for the conformal factor and those of the metric, in these theories, are constrained by the topological density.  For such classical theories, a nonlocal formulation is possible, that differs significantly from the local one(s) presented in the recent literature. This is obtained by including the same finite renormalization of the topological density discussed in \cite{Mazur:2001aa}. The use of these actions in a cosmological context provides an important starting point for the analysis of nonlocal cosmological models, that induce significant modifications in the evolution of gravitational waves in the early universe. \\
\centerline{\bf Acknowledgements} 
We thank Mario Cret\`i, Stefano Lionetti and Riccardo Tommasi for discussions. This work is partially supported by INFN within the Iniziativa Specifica QFT-HEP.  M. M. M. is supported by the European Research Council (ERC) under the European Union as Horizon 2020 research and innovation program (grant agreement No818066) and by Deutsche Forschungsgemeinschaft (DFG, German Research Foundation) under Germany's Excellence Strategy EXC-2181/1 - 390900948 (the Heidelberg STRUCTURES Cluster of Excellence).

\appendix
\section{Appendix}

 In this appendix we define our conventions and study the behaviour of various tensors under a Weyl transformation. 
 We define 
 \beq
R^\lambda_{\,\,\mu\sigma \nu}=\partial_\sigma \Gamma^\lambda_{\mu\nu} -\partial_\nu\Gamma^\lambda_{\mu\sigma} +
\Gamma^\lambda_{\rho\sigma}\Gamma^\rho_{\mu\nu} -\Gamma^\lambda_{\rho\nu}\Gamma^\rho_{\mu\sigma}.
\eeq
We choose a fiducial metric as in \eqref{mar}. The Christoffel symbol then transforms as
\beq
\label{dg}
\Gamma_{\mu\nu}^\lambda=\bar \Gamma_{\mu\nu}^\lambda+\de_\mu^\lambda\bar \nabla_\nu\phi+\de^\lambda_\nu\bar \nabla_\mu\phi-\bar g_{\mu\nu}\bar \nabla^\lambda\phi,
\eeq
and the curvature tensors
\bea
R^\lambda{}_{\mu\sigma\nu} &=& \bar R^\lambda{}_{\mu\sigma\nu}+\de^\lambda_\nu\bar \Delta_{\mu\sigma}- \de^\lambda_\sigma\bar \Delta_{\mu\nu}+\bar g_{\mu\sigma}\bar \Delta^\lambda_\nu-\bar g_{\mu\nu}\bar \nabla^\lambda_\sigma +
(\de^\lambda_\nu\bar g_{\mu\sigma}- \de^\lambda_\sigma\bar g_{\mu\nu})\bar \nabla_\rho\phi\bar \nabla^\rho\phi, \\
\label{Rmunu underr finite}
R_{\mu\nu}&=&\bar R_{\mu\nu}-(d-2)\bar \Delta_{\mu\nu}-\bar g_{\mu\nu}[\bar \square \phi+(d-2)\bar \nabla_\lambda\phi\bar \nabla^\lambda\phi], \\
\label{R under finite Weyl}
R&=&e^{-2\phi}[\bar R-2(d-1)\bar \square \phi -(d-1)(d-2)\bar \nabla_\lambda\phi\bar \nabla^\lambda\phi], \eea
where we have defined the symmetric tensor $\bar \Delta_{\mu \nu}$ and its trace $\bar \Delta$ as
\beq \bar \Delta_{\mu \nu} = \bnabla_\mu \bnabla_\nu \phi - \bnabla_\mu \phi \bnabla_\nu \phi , \qquad \bar \Delta=\bar g^{\mu \nu} \bar \Delta_{\mu \nu} =\bar \square \phi - \bnabla_\lambda \phi \bnabla^\lambda \phi. \label{dd}\eeq
Notice that the symmetry of this tensor is ensured by the relation
\beq 
\bnabla_\mu \bnabla_\nu \phi = (\bpde_\mu \bpde_\nu - \bar \Gamma^\lambda_{\mu \nu} \bpde_\lambda) \phi= 
( \bpde_\nu \bpde_\mu - \bar\Gamma^\lambda_{\nu \mu} \bpde_\lambda )\phi = \bnabla_\nu \bnabla_\mu \phi . \eeq
In order to find $E$, we need the square of the curvatures
\bea R_{\mu \nu \rho \sigma}^2 &=& e^{-4\phi} \lt
\bar R_{\mu \nu \rho \sigma}^2 - 8 \bar R^{\mu \nu} \bar \Delta_{\mu \nu} - 4 \bar R\bnabla_\lambda \phi \bnabla^\lambda \phi + 4(d-2)\bar \Delta_{\mu \nu}^2 + 4\bar \Delta ^2  \right. \nn && \left.
+ 8(d-1)\bar \Delta \bnabla_\lambda \phi \bnabla^\lambda \phi  + 2d(d-1)(\bnabla_\lambda \phi \bnabla^\lambda \phi)^2 \rt 
\\
R_{\mu \nu}^2 &=& e^{-4 \phi} \lq\bar R_{\mu \nu}^2 - 2(d-2) \bar R^{\mu \nu} \bar\Delta_{\mu \nu} 
- 2\bar R \bar\square \phi - 2(d-2) \bar R (\bnabla \phi)^2 + (d-2)^2 (\bnabla_\mu \bnabla_\nu \phi )^2 \right. \nn && \left.
- 2(d-2)^2 \bnabla_\mu \bnabla_ \nu \phi \bnabla^\mu \phi \bnabla^\nu \phi
+ (3d-4) (\bar\square \phi)^2  \right. \nn && \left. + 2(d-2)(2d-3)\bar\square \phi \bnabla_\lambda \phi \bnabla^\lambda \phi + (d-1)(d-2)^2 (\bnabla_\lambda \phi \bnabla^\lambda \phi)^2 \rq
\\
 R^2 &=& e^{-4 \phi} \lq \bar R^2 - 4(d-1) \bar R \bar \square \phi - 2(d-1)(d-2)\bar R\bnabla_\lambda \phi \bnabla^\lambda \phi
+ 4 (d-1)^2 (\bar \square \phi)^2 \right. \nn && \left. + 4(d-1)^2 (d-2) \bar \square \phi \bnabla_\lambda \phi \bnabla^\lambda \phi
+ (d-1)^2 (d-2)^2 (\bnabla_\lambda \phi \bnabla^\lambda \phi)^2
 \rq .\eea
By using these relations, we can rewrite the rescaled $E$ as
\bea E&=& \bar E + 8 (d-3)\bar R^{\mu \nu} \bar\Delta_{\mu \nu} - 2(d-4)(d-3) \bar R (\bnabla \phi)^2 - 4 (d-3) \bar R \bar \square \phi  - 4(d-3)(d-2) (\bnabla_\mu \bnabla_\nu \phi)^2 \nn && + 8 (d-2)(d-3) \bnabla_\mu \bnabla_\nu \phi \bnabla^\mu \phi \bnabla^\nu \phi 
+ 4(d-3)(d-2)(\bar \square \phi)^2 + 4(d-3)^2 (d-2)\bar\square \phi (\bnabla \phi)^2 \nn && + (d-4)(d-3)(d-2)(d-1)(\bnabla \phi)^4 . \eea
The tensor relations for $\bar \Delta_{\mu \nu}$
$$\bar\Delta^2=(\bar\Box\phi)^2+(\bnabla_\lambda \phi \bnabla^\lambda \phi)^2-2\bnabla_\lambda \phi\bnabla^\lambda \phi\bar\Box\phi,$$
$$ \bar R_{\mu\nu}\bar \Delta^{\mu\nu}=\bar R_{\mu\nu}\bar\nabla^\mu\bar \nabla^\nu\phi-\bar R_{\mu\nu}\bar\nabla^\mu\phi\bar\nabla^\nu\phi,
$$
$$
\bar \Delta_{\mu\nu}^2=(\bar\nabla_\mu\bar\nabla_\nu\phi)^2+(\bnabla_\lambda \phi\bnabla^\lambda \phi)^2-2(\bar\nabla_\mu\bar\nabla_\nu\phi\bar\nabla^\mu\phi\bar\nabla^\nu\phi),
$$
and
\beqa
\nabla_c\nabla_b V_a-\nabla_b\nabla_c V_a &=& R^d_{\,\, a b c} V_d \nn\\
\nabla_\mu\nabla_\rho\nabla_\sigma\phi -\nabla_\rho\nabla_\mu\nabla_\sigma\phi &=&
R^\epsilon_{\,\,\sigma\rho\mu}\nabla_\epsilon \phi \nn\\
\nabla_\mu R^{\mu\nu}=\frac{1}{2}\nabla^\nu R.
\eeqa
have been used for the derivation of the rescaled expressions of $E$.\\

As a final remark, we recall that a functional differentiation generates covariant derivatives of delta functions, which are scalar densities, that can be rewritten as ordinary scalars with the trick $\delta(x)\to (\delta(x)/\sqrt{g}) \sqrt{g}$, where 
$\delta(x)/\sqrt{g}$ is an ordinary scalar. At this stage one can apply, as usual, the Liebnitz rule on  tensor products, as well as integration by parts, using for the scalar density 
$\sqrt{g}$ the relation 
\beq
\nabla_\mu \sqrt{g}=\left[\partial_\mu -\left(\frac{1}{\sqrt{g}}\partial_\mu \sqrt{g}\right)\right] \sqrt{g}=0.
\eeq
\section{Functional relations and boundary terms in $V_{C^2}$}
We summarize the following expressions for the Weyl squared terms in 4 and in $d$ dimensions

\begin{align}
(C^{(4)})^2&=R^{\mu \nu \rho \sigma}R_{\mu \nu \rho \sigma}-2R^{\mu \nu }R_{\mu \nu}+R^2\notag \\ 
(C^{(d)})^2&=R^{\mu \nu \rho \sigma}R_{\mu \nu \rho \sigma}-\frac{4}{d-2}R^{\mu \nu }R_{\mu \nu}+\frac{2}{(d-2)(d-1)}R^2\notag \\ 
(C^{(d)})^2&=(C^{(4)})^2+\frac{d-4}{d-2}\left(2R^{\mu \nu }R_{\mu \nu}-\frac{d+1}{3(d-1)}R^2\right).\notag 
\end{align}
We have the following relation 
\begin{equation}
2g_{\mu \nu}\frac{\delta}{\sqrt{-g}\delta g_{\mu \nu}}\int d^d x  \sqrt{-g} (C^{(d)})^2=(d-4)(C^{(d)})^2.
\end{equation}
By using the relation between  $(C^{(4)})^2$ and $(C^{(d)})^2$ we can write 
\begin{equation}
2g_{\mu \nu}\frac{\delta}{\sqrt{-g}\delta g_{\mu \nu}}\int d^d x  \sqrt{-g} \left((C^{(4)})^2+\frac{d-4}{d-2}\left(\frac{d-4}{d-2}R^{\mu \nu }R_{\mu \nu}-\frac{d+1}{3(d-1)}R^2\right)\right)=(d-4)(C^{(d)})^2.
\end{equation}
By rearranging the terms above we get
\beqa
\label{var}
2g_{\mu \nu}\frac{\delta}{\sqrt{-g}\delta g_{\mu \nu}}\int d^d x  \sqrt{-g} (C^{(4)})^2 &=&(d-4)(C^{(d)})^2\nn
&& -2g_{\mu \nu}\frac{\delta}{\sqrt{-g}\delta g_{\mu \nu}}\int d^d x \sqrt{-g}\frac{d-4}{d-2}\left(2R^{\mu \nu }R_{\mu \nu}-\frac{d+1}{3(d-1)}R^2\right).\nn
\eeqa
By a direct computation we obtain
\beqa
&& 2g_{\mu \nu}\frac{\delta}{\sqrt{-g}\delta g_{\mu \nu}}\int d^d x \sqrt{-g}\frac{d-4}{d-2}\left(2R^{\mu \nu }R_{\mu \nu}-\frac{d+1}{3(d-1)}R^2\right)\nn
&& \qquad \qquad \qquad =(d-4)\left\lbrace \frac{d-4}{d-2}\left(2R^{\mu \nu }R_{\mu \nu}-\frac{d+1}{3(d-1)}R^2\right)-\frac{2}{3}\Box R \right\rbrace.
\eeqa
Then we can substitute this expression in \eqref{var} to obtain
\begin{align}\label{finalvar}
2g_{\mu \nu}\frac{\delta}{\sqrt{-g}\delta g_{\mu \nu}}\int d^d x  \sqrt{-g} (C^{(4)})^2&=(d-4)\left[(C^{(d)})^2+\frac{2}{3}\Box R \right] -\frac{(d-4)^2}{(d-2)}\left(2R^{\mu \nu }R_{\mu \nu}-\frac{d+1}{3(d-1)}R^2\right) \notag \\
&=(d-4)\left[(C^{(4)})^2+\frac{2}{3}\Box R+\frac{(d-4)}{(d-2)}\left(2R^{\mu \nu }R_{\mu \nu}-\frac{d+1}{3(d-1)}R^2\right) \right]\notag \\&-\frac{(d-4)^2}{(d-2)}\left(2R^{\mu \nu }R_{\mu \nu}-\frac{d+1}{3(d-1)}R^2\right)\notag\\
&=(d-4)\left[(C^{(4)})^2+\frac{2}{3}\Box R \right].
\end{align}
We obtain the same result by a direct computation
\begin{equation}
2g_{\mu \nu}\frac{\delta}{\sqrt{-g}\delta g_{\mu \nu}}\int d^d x  \sqrt{-g} (C^{(4)})^2=(d-4)\left[(C^{(4)})^2+\frac{2}{3}\Box R \right].
\end{equation}

\section{Consistency of the expansion and the $d\to$ 4 limit} 
The consistency between the functional differentiation and the $d\to 4$ limit of the effective action can be shown as follows.  From 
\beq
V_E(g,d)-V_E(\bar g,d)=\int d^d x \sqrt{{\bar g}} e^{(d-4)\phi}\left[ 
\bar E +(d-3) \bar\nabla_\mu J^\mu +(d-3)(d-4) K\right] -\int d^d x \sqrt{\bar g } \bar E
\eeq
we obtain
\begin{align}
\lim_{d\to 4}\left[\frac{\delta}{\delta\phi}\frac{1}{d-4}\left(V_E(g,d)-V_E(\bar g,d)\right)\right]=& \notag\\
 \lim_{d\to 4} 
\left(\int d^d x \sqrt{g}E\delta_{x y} 
 +\frac{d-3}{d-4}\int d^d x \sqrt{\bar g}\bar{\nabla}_\mu\left[ 
e^{(d-4)\phi}\left(\frac{\delta J^\mu}{\delta\phi} + 4 (d-2)(d-4)\bar{\nabla}_\mu\phi\bar{\nabla}_\nu\phi\bar{\nabla}^\nu \delta_{x y}\right)\right]\right)&&\nn
=\int d^d x \sqrt{g}E\delta_{x y} +\int d^d x \sqrt{\bar{g}}\bar{\nabla}_\mu\left[\phi \frac{\delta J^\mu}{\delta \phi} + 
8 \bar{\nabla}^\mu\phi \bar{\nabla}_\nu \phi \bar{\nabla}^\nu\delta_{x y}\right] &.   
\label{oone}
\end{align}
On the other end we have 

\begin{align}
\lim_{d\to 4}\left[\frac{1}{d-4}\left(V_E(g,d)-V_E(\bar g,d)\right)\right]&=\lim_{d\to 4}\left[\frac{d-3}{d-4}\int d^d x \sqrt{\bar g}
\bar{\nabla}_\mu J^\mu +\int d^d x \sqrt{\bar{g}}\left[ \phi \bar{E} +(d-3)\phi \bar{\nabla}_\mu J^\mu +(d-3) K\right]\right].
\end{align}
Notice that the first term on the rhs of the equation above is of the form $0/0$, and can be neglected under the assumption that the $d\to 4$ limit is performed after removing the boundary contribution.  
Differentiating the expression above we obtain 
\begin{align}
\frac{\delta}{\delta \phi}\lim_{d\to 4}\left[\frac{1}{d-4}\left(V_E(g,d)-V_E(\bar g,d)\right)\right]&= \int d^d x \sqrt{\bar{g}}\bar E \delta_{x y} +\int d^d x \sqrt{\bar{g}}\left[ \phi \bar{\nabla}_\mu \frac{\delta J^\mu}{\delta \phi}+\frac{\delta K}{\delta\phi}\right]\nn 
&=\int d^d x \sqrt{\bar g}\left( \bar E +\bar{\nabla}_\mu J^\mu\right)\delta_{x y}\nn 
&+ 
\int d^d x \sqrt{\bar g}\bar{\nabla}_\mu\left[ \phi\frac{\delta J^\mu}{\delta\phi}\right] 
+ \int d^d x \sqrt{\bar g}\left[ \frac{\delta K}{\delta\phi }-\nabla_\mu \phi \frac{\delta J^\mu}{\delta\phi}\right]. 
\end{align}
Now, using 
\beq
 \frac{\delta K}{\delta\phi }-\nabla_\mu \phi \frac{\delta J^\mu}{\delta\phi}=8\bar \nabla^\mu\left(\bar \nabla_\mu \phi \bar\nabla_\nu \delta_{x y} \bar \nabla^\nu\phi\right)
\eeq
one can show the agreement with \eqref{oone}.


\begin{thebibliography}{10}

\bibitem{Charmousis:2014mia}
C.~Charmousis, {\it {From Lovelock to Horndeski`s Generalized Scalar Tensor
  Theory}},  {\em Lect. Notes Phys.} {\bf 892} (2015) 25--56,
  [\href{http://xxx.lanl.gov/abs/1405.1612}{{\tt arXiv:1405.1612}}].

\bibitem{Starobinsky:1980te}
A.~A. Starobinsky, {\it {A new type of isotropic cosmological models without
  singularity}},  {\em Phys. Lett.} {\bf B91} (1980) 99--102.

\bibitem{Antoniadis:2011ib}
I.~Antoniadis, P.~O. Mazur, and E.~Mottola, {\it {Conformal Invariance, Dark
  Energy, and CMB Non-Gaussianity}},  {\em JCAP} {\bf 1209} (2012) 024,
  [\href{http://xxx.lanl.gov/abs/1103.4164}{{\tt arXiv:1103.4164}}].

\bibitem{Hamada:2014pba}
K.-j. Hamada, {\it {Renormalization group analysis for quantum gravity with a
  single dimensionless coupling}},  {\em Phys. Rev. D} {\bf 90} (2014), no.~8
  084038, [\href{http://xxx.lanl.gov/abs/1407.4532}{{\tt arXiv:1407.4532}}].

\bibitem{Visser:2002ew}
M.~Visser, {\it {Sakharov's induced gravity: A Modern perspective}},  {\em Mod.
  Phys. Lett. A} {\bf 17} (2002) 977--992,
  [\href{http://xxx.lanl.gov/abs/gr-qc/0204062}{{\tt gr-qc/0204062}}].

\bibitem{Codello:2012sn}
A.~Codello, G.~D'Odorico, C.~Pagani, and R.~Percacci, {\it {The Renormalization
  Group and Weyl-invariance}},  {\em Class.Quant.Grav.} {\bf 30} (2013) 115015,
  [\href{http://xxx.lanl.gov/abs/1210.3284}{{\tt arXiv:1210.3284}}].

\bibitem{Coriano:2013xua}
C.~Corian\`o, L.~Delle~Rose, C.~Marzo, and M.~Serino, {\it {Conformal Trace
  Relations from the Dilaton Wess-Zumino Action}},  {\em Phys. Lett. B} {\bf
  726} (2013), no.~4-5 896--905, [\href{http://xxx.lanl.gov/abs/1306.4248}{{\tt
  arXiv:1306.4248}}].

\bibitem{Coriano:2013nja}
C.~Corian\`o, L.~Delle~Rose, C.~Marzo, and M.~Serino, {\it {The dilaton
  Wess-Zumino action in six dimensions from Weyl gauging: local anomalies and
  trace relations}},  {\em Class. Quant. Grav.} {\bf 31} (2014) 105009,
  [\href{http://xxx.lanl.gov/abs/1311.1804}{{\tt arXiv:1311.1804}}].

\bibitem{tHooft:2016uxd}
G.~'t~Hooft, {\it {Local conformal symmetry in black holes, standard model, and
  quantum gravity}},  {\em Int. J. Mod. Phys. D} {\bf 26} (2016), no.~03
  1730006.

\bibitem{Armillis:2011hj}
R.~Armillis, C.~Corian\`o, L.~Delle~Rose, and A.~R. Fazio, {\it {Comments on
  Anomaly Cancellations by Pole Subtractions and Ghost Instabilities with
  Gravity}},  {\em Class. Quant. Grav.} {\bf 28} (2011) 145004,
  [\href{http://xxx.lanl.gov/abs/1103.1590}{{\tt arXiv:1103.1590}}].

\bibitem{Coriano:2005js}
C.~Corian\`o, N.~Irges, and E.~Kiritsis, {\it {On the effective theory of low
  scale orientifold string vacua}},  {\em Nucl. Phys.} {\bf B746} (2006)
  77--135, [\href{http://xxx.lanl.gov/abs/hep-ph/0510332}{{\tt
  hep-ph/0510332}}].

\bibitem{Mann:1992ar}
R.~B. Mann and S.~F. Ross, {\it {The D ---\ensuremath{>} 2 limit of general
  relativity}},  {\em Class. Quant. Grav.} {\bf 10} (1993) 1405--1408,
  [\href{http://xxx.lanl.gov/abs/gr-qc/9208004}{{\tt gr-qc/9208004}}].

\bibitem{Anastasiou:2020zwc}
G.~Anastasiou, O.~Miskovic, R.~Olea, and I.~Papadimitriou, {\it {Counterterms,
  Kounterterms, and the variational problem in AdS gravity}},  {\em JHEP} {\bf
  08} (2020) 061, [\href{http://xxx.lanl.gov/abs/2003.0642}{{\tt
  arXiv:2003.0642}}].

\bibitem{Matsumoto:2022fln}
M.~Matsumoto and Y.~Nakayama, {\it {Dilaton invading from infinitesimal extra
  dimension}},  \href{http://xxx.lanl.gov/abs/2202.1353}{{\tt
  arXiv:2202.1353}}.

\bibitem{1984PhLB..134...56R}
R.~J. {Riegert}, {\it {A non-local action for the trace anomaly}},  {\em
  Physics Letters B} {\bf 134} (Jan., 1984) 56--60.

\bibitem{Mazur:2001aa}
P.~O. Mazur and E.~Mottola, {\it {Weyl cohomology and the effective action for
  conformal anomalies}},  {\em Phys.Rev.} {\bf D64} (2001) 104022,
  [\href{http://xxx.lanl.gov/abs/hep-th/0106151}{{\tt hep-th/0106151}}].

\bibitem{Coriano:2022knl}
C.~Corian\`o and M.~M. Maglio, {\it {Einstein Gauss-Bonnet Theories as
  Ordinary, Wess-Zumino Conformal Anomaly Actions}},
  \href{http://xxx.lanl.gov/abs/2201.0751}{{\tt arXiv:2201.0751}}.

\bibitem{Giannotti:2008cv}
M.~Giannotti and E.~Mottola, {\it {The Trace Anomaly and Massless Scalar
  Degrees of Freedom in Gravity}},  {\em Phys. Rev.} {\bf D79} (2009) 045014,
  [\href{http://xxx.lanl.gov/abs/0812.0351}{{\tt arXiv:0812.0351}}].

\bibitem{Coriano:2018bsy}
C.~Corian\`o and M.~M. Maglio, {\it {The general 3-graviton vertex ($TTT$) of
  conformal field theories in momentum space in $d =4$}},  {\em Nucl. Phys.}
  {\bf B937} (2018) 56--134, [\href{http://xxx.lanl.gov/abs/1808.1022}{{\tt
  arXiv:1808.1022}}].

\bibitem{Armillis:2009pq}
R.~Armillis, C.~Corian\`{o}, and L.~Delle~Rose, {\it {Conformal Anomalies and
  the Gravitational Effective Action: The $TJJ$ Correlator for a Dirac
  Fermion}},  {\em Phys. Rev.} {\bf D81} (2010) 085001,
  [\href{http://xxx.lanl.gov/abs/0910.3381}{{\tt arXiv:0910.3381}}].

\bibitem{Glavan:2019inb}
D.~Glavan and C.~Lin, {\it {Einstein-Gauss-Bonnet Gravity in Four-Dimensional
  Spacetime}},  {\em Phys. Rev. Lett.} {\bf 124} (2020), no.~8 081301,
  [\href{http://xxx.lanl.gov/abs/1905.0360}{{\tt arXiv:1905.0360}}].

\bibitem{Lovelock:1971yv}
D.~Lovelock, {\it {The Einstein tensor and its generalizations}},  {\em J.
  Math. Phys.} {\bf 12} (1971) 498--501.

\bibitem{Gurses:2020ofy}
G.~Metin, T.~C. Sisman, and T.~Bayram, {\it {Is there a novel
  Einstein\textendash{}Gauss\textendash{}Bonnet theory in four dimensions?}},
  {\em Eur. Phys. J. C} {\bf 80} (2020), no.~7 647,
  [\href{http://xxx.lanl.gov/abs/2004.0339}{{\tt arXiv:2004.0339}}].

\bibitem{Hennigar:2020lsl}
R.~A. Hennigar, D.~Kubiz\v{n}\'ak, R.~B. Mann, and C.~Pollack, {\it {On taking
  the D to 4 limit of Gauss-Bonnet gravity: theory and solutions}},  {\em JHEP}
  {\bf 07} (2020) 027, [\href{http://xxx.lanl.gov/abs/2004.0947}{{\tt
  arXiv:2004.0947}}].

\bibitem{Fernandes:2020nbq}
P.~G.~S. Fernandes, P.~Carrilho, T.~Clifton, and D.~J. Mulryne, {\it
  {Derivation of Regularized Field Equations for the Einstein-Gauss-Bonnet
  Theory in Four Dimensions}},  {\em Phys. Rev. D} {\bf 102} (2020), no.~2
  024025, [\href{http://xxx.lanl.gov/abs/2004.0836}{{\tt arXiv:2004.0836}}].

\bibitem{Lu:2020iav}
H.~Lu and Y.~Pang, {\it {Horndeski gravity as $D \rightarrow 4$ limit of
  Gauss-Bonnet}},  {\em Phys. Lett. B} {\bf 809} (2020) 135717,
  [\href{http://xxx.lanl.gov/abs/2003.1155}{{\tt arXiv:2003.1155}}].

\bibitem{Easson:2020mpq}
D.~A. Easson, T.~Manton, and A.~Svesko, {\it {$D\to4$ Einstein-Gauss-Bonnet
  gravity and beyond}},  {\em JCAP} {\bf 10} (2020) 026,
  [\href{http://xxx.lanl.gov/abs/2005.1229}{{\tt arXiv:2005.1229}}].

\bibitem{Kobayashi:2020wqy}
T.~Kobayashi, {\it {Effective scalar-tensor description of regularized Lovelock
  gravity in four dimensions}},  {\em JCAP} {\bf 07} (2020) 013,
  [\href{http://xxx.lanl.gov/abs/2003.1277}{{\tt arXiv:2003.1277}}].

\bibitem{Konoplya:2020qqh}
R.~A. Konoplya and A.~Zhidenko, {\it {Black holes in the four-dimensional
  Einstein-Lovelock gravity}},  {\em Phys. Rev. D} {\bf 101} (2020), no.~8
  084038, [\href{http://xxx.lanl.gov/abs/2003.0778}{{\tt arXiv:2003.0778}}].

\bibitem{Bonifacio:2020vbk}
J.~Bonifacio, K.~Hinterbichler, and L.~A. Johnson, {\it {Amplitudes and 4D
  Gauss-Bonnet Theory}},  {\em Phys. Rev. D} {\bf 102} (2020), no.~2 024029,
  [\href{http://xxx.lanl.gov/abs/2004.1071}{{\tt arXiv:2004.1071}}].

\bibitem{Ai:2020peo}
W.-Y. Ai, {\it {A note on the novel 4D
  Einstein\textendash{}Gauss\textendash{}Bonnet gravity}},  {\em Commun. Theor.
  Phys.} {\bf 72} (2020), no.~9 095402,
  [\href{http://xxx.lanl.gov/abs/2004.0285}{{\tt arXiv:2004.0285}}].

\bibitem{Wei:2020ght}
S.-W. Wei and Y.-X. Liu, {\it {Testing the nature of Gauss-Bonnet gravity by
  four-dimensional rotating black hole shadow}},  {\em Eur. Phys. J. Plus} {\bf
  136} (2021), no.~4 436, [\href{http://xxx.lanl.gov/abs/2003.0776}{{\tt
  arXiv:2003.0776}}].

\bibitem{Aoki:2020lig}
K.~Aoki, M.~A. Gorji, and S.~Mukohyama, {\it {A consistent theory of $D \to 4$
  Einstein-Gauss-Bonnet gravity}},  {\em Phys. Lett. B} {\bf 810} (2020)
  135843, [\href{http://xxx.lanl.gov/abs/2005.0385}{{\tt arXiv:2005.0385}}].

\bibitem{Nojiri:2020tph}
S.~Nojiri and S.~D. Odintsov, {\it {Novel cosmological and black hole solutions
  in Einstein and higher-derivative gravity in two dimensions}},  {\em EPL}
  {\bf 130} (2020), no.~1 10004, [\href{http://xxx.lanl.gov/abs/2004.0140}{{\tt
  arXiv:2004.0140}}].

\bibitem{Konoplya:2020bxa}
R.~A. Konoplya and A.~F. Zinhailo, {\it {Quasinormal modes, stability and
  shadows of a black hole in the 4D
  Einstein\textendash{}Gauss\textendash{}Bonnet gravity}},  {\em Eur. Phys. J.
  C} {\bf 80} (2020), no.~11 1049,
  [\href{http://xxx.lanl.gov/abs/2003.0118}{{\tt arXiv:2003.0118}}].

\bibitem{Guo:2020zmf}
M.~Guo and P.-C. Li, {\it {Innermost stable circular orbit and shadow of the
  $4D$ Einstein\textendash{}Gauss\textendash{}Bonnet black hole}},  {\em Eur.
  Phys. J. C} {\bf 80} (2020), no.~6 588,
  [\href{http://xxx.lanl.gov/abs/2003.0252}{{\tt arXiv:2003.0252}}].

\bibitem{Fernandes:2020rpa}
P.~G.~S. Fernandes, {\it {Charged black holes in AdS spaces in 4D Einstein
  Gauss-Bonnet gravity}},  {\em Phys. Lett. B} {\bf 805} (2020) 135468,
  [\href{http://xxx.lanl.gov/abs/2003.0549}{{\tt arXiv:2003.0549}}].

\bibitem{Casalino:2020kbt}
A.~Casalino, A.~Colleaux, M.~Rinaldi, and S.~Vicentini, {\it {Regularized
  Lovelock gravity}},  {\em Phys. Dark Univ.} {\bf 31} (2021) 100770,
  [\href{http://xxx.lanl.gov/abs/2003.0706}{{\tt arXiv:2003.0706}}].

\bibitem{Hegde:2020xlv}
K.~Hegde, A.~Naveena~Kumara, C.~L.~A. Rizwan, A.~K. M., and M.~S. Ali, {\it
  {Thermodynamics, Phase Transition and Joule Thomson Expansion of novel 4-D
  Gauss Bonnet AdS Black Hole}},  \href{http://xxx.lanl.gov/abs/2003.0877}{{\tt
  arXiv:2003.0877}}.

\bibitem{Ghosh:2020vpc}
S.~G. Ghosh and S.~D. Maharaj, {\it {Radiating black holes in the novel 4D
  Einstein\textendash{}Gauss\textendash{}Bonnet gravity}},  {\em Phys. Dark
  Univ.} {\bf 30} (2020) 100687, [\href{http://xxx.lanl.gov/abs/2003.0984}{{\tt
  arXiv:2003.0984}}].

\bibitem{Doneva:2020ped}
D.~D. Doneva and S.~S. Yazadjiev, {\it {Relativistic stars in 4D
  Einstein-Gauss-Bonnet gravity}},  {\em JCAP} {\bf 05} (2021) 024,
  [\href{http://xxx.lanl.gov/abs/2003.1028}{{\tt arXiv:2003.1028}}].

\bibitem{Zhang:2020qew}
Y.-P. Zhang, S.-W. Wei, and Y.-X. Liu, {\it {Spinning Test Particle in
  Four-Dimensional Einstein\textendash{}Gauss\textendash{}Bonnet Black Holes}},
   {\em Universe} {\bf 6} (2020), no.~8 103,
  [\href{http://xxx.lanl.gov/abs/2003.1096}{{\tt arXiv:2003.1096}}].

\bibitem{Konoplya:2020ibi}
R.~A. Konoplya and A.~Zhidenko, {\it {BTZ black holes with higher curvature
  corrections in the 3D Einstein-Lovelock gravity}},  {\em Phys. Rev. D} {\bf
  102} (2020), no.~6 064004, [\href{http://xxx.lanl.gov/abs/2003.1217}{{\tt
  arXiv:2003.1217}}].

\bibitem{Singh:2020xju}
D.~V. Singh and S.~Siwach, {\it {Thermodynamics and P-v criticality of
  Bardeen-AdS Black Hole in 4$D$ Einstein-Gauss-Bonnet Gravity}},  {\em Phys.
  Lett. B} {\bf 808} (2020) 135658,
  [\href{http://xxx.lanl.gov/abs/2003.1175}{{\tt arXiv:2003.1175}}].

\bibitem{Ghosh:2020syx}
S.~G. Ghosh and R.~Kumar, {\it {Generating black holes in $4D$
  Einstein-Gauss-Bonnet gravity}},  {\em Class. Quant. Grav.} {\bf 37} (2020),
  no.~24 245008, [\href{http://xxx.lanl.gov/abs/2003.1229}{{\tt
  arXiv:2003.1229}}].

\bibitem{Konoplya:2020juj}
R.~A. Konoplya and A.~Zhidenko, {\it {(In)stability of black holes in the $4D$
  Einstein\textendash{}Gauss\textendash{}Bonnet and
  Einstein\textendash{}Lovelock gravities}},  {\em Phys. Dark Univ.} {\bf 30}
  (2020) 100697, [\href{http://xxx.lanl.gov/abs/2003.1249}{{\tt
  arXiv:2003.1249}}].

\bibitem{Kumar:2020uyz}
A.~Kumar and R.~Kumar, {\it {Bardeen black holes in the novel $4D$
  Einstein-Gauss-Bonnet gravity}},
  \href{http://xxx.lanl.gov/abs/2003.1310}{{\tt arXiv:2003.1310}}.

\bibitem{Zhang:2020qam}
C.-Y. Zhang, P.-C. Li, and M.~Guo, {\it {Greybody factor and power spectra of
  the Hawking radiation in the $4D$
  Einstein\textendash{}Gauss\textendash{}Bonnet de-Sitter gravity}},  {\em Eur.
  Phys. J. C} {\bf 80} (2020), no.~9 874,
  [\href{http://xxx.lanl.gov/abs/2003.1306}{{\tt arXiv:2003.1306}}].

\bibitem{HosseiniMansoori:2020yfj}
S.~A. Hosseini~Mansoori, {\it {Thermodynamic geometry of the novel 4-D
  Gauss\textendash{}Bonnet AdS black hole}},  {\em Phys. Dark Univ.} {\bf 31}
  (2021) 100776, [\href{http://xxx.lanl.gov/abs/2003.1338}{{\tt
  arXiv:2003.1338}}].

\bibitem{Wei:2020poh}
S.-W. Wei and Y.-X. Liu, {\it {Extended thermodynamics and microstructures of
  four-dimensional charged Gauss-Bonnet black hole in AdS space}},  {\em Phys.
  Rev. D} {\bf 101} (2020), no.~10 104018,
  [\href{http://xxx.lanl.gov/abs/2003.1427}{{\tt arXiv:2003.1427}}].

\bibitem{Singh:2020nwo}
D.~V. Singh, S.~G. Ghosh, and S.~D. Maharaj, {\it {Clouds of strings in 4$D$
  Einstein\textendash{}Gauss\textendash{}Bonnet black holes}},  {\em Phys. Dark
  Univ.} {\bf 30} (2020) 100730, [\href{http://xxx.lanl.gov/abs/2003.1413}{{\tt
  arXiv:2003.1413}}].

\bibitem{Churilova:2020aca}
M.~S. Churilova, {\it {Quasinormal modes of the Dirac field in the consistent
  4D Einstein\textendash{}Gauss\textendash{}Bonnet gravity}},  {\em Phys. Dark
  Univ.} {\bf 31} (2021) 100748, [\href{http://xxx.lanl.gov/abs/2004.0051}{{\tt
  arXiv:2004.0051}}].

\bibitem{Islam:2020xmy}
S.~U. Islam, R.~Kumar, and S.~G. Ghosh, {\it {Gravitational lensing by black
  holes in the $4D$ Einstein-Gauss-Bonnet gravity}},  {\em JCAP} {\bf 09}
  (2020) 030, [\href{http://xxx.lanl.gov/abs/2004.0103}{{\tt
  arXiv:2004.0103}}].

\bibitem{Mishra:2020gce}
A.~K. Mishra, {\it {Quasinormal modes and strong cosmic censorship in the
  regularised 4D Einstein\textendash{}Gauss\textendash{}Bonnet gravity}},  {\em
  Gen. Rel. Grav.} {\bf 52} (2020), no.~11 106,
  [\href{http://xxx.lanl.gov/abs/2004.0124}{{\tt arXiv:2004.0124}}].

\bibitem{Konoplya:2020cbv}
R.~A. Konoplya and A.~F. Zinhailo, {\it {Grey-body factors and Hawking
  radiation of black holes in $4D$ Einstein-Gauss-Bonnet gravity}},  {\em Phys.
  Lett. B} {\bf 810} (2020) 135793,
  [\href{http://xxx.lanl.gov/abs/2004.0224}{{\tt arXiv:2004.0224}}].

\bibitem{Zhang:2020sjh}
C.-Y. Zhang, S.-J. Zhang, P.-C. Li, and M.~Guo, {\it {Superradiance and
  stability of the regularized 4D charged Einstein-Gauss-Bonnet black hole}},
  {\em JHEP} {\bf 08} (2020) 105,
  [\href{http://xxx.lanl.gov/abs/2004.0314}{{\tt arXiv:2004.0314}}].

\bibitem{EslamPanah:2020hoj}
B.~Eslam~Panah, K.~Jafarzade, and S.~H. Hendi, {\it {Charged 4D
  Einstein-Gauss-Bonnet-AdS black holes: Shadow, energy emission, deflection
  angle and heat engine}},  {\em Nucl. Phys. B} {\bf 961} (2020) 115269,
  [\href{http://xxx.lanl.gov/abs/2004.0405}{{\tt arXiv:2004.0405}}].

\bibitem{Aragon:2020qdc}
A.~Arag\'on, R.~B\'ecar, P.~A. Gonz\'alez, and Y.~V\'asquez, {\it {Perturbative
  and nonperturbative quasinormal modes of 4D
  Einstein\textendash{}Gauss\textendash{}Bonnet black holes}},  {\em Eur. Phys.
  J. C} {\bf 80} (2020), no.~8 773,
  [\href{http://xxx.lanl.gov/abs/2004.0563}{{\tt arXiv:2004.0563}}].

\bibitem{Aoki:2020iwm}
K.~Aoki, M.~A. Gorji, and S.~Mukohyama, {\it {Cosmology and gravitational waves
  in consistent $D\to 4$ Einstein-Gauss-Bonnet gravity}},  {\em JCAP} {\bf 09}
  (2020) 014, [\href{http://xxx.lanl.gov/abs/2005.0842}{{\tt
  arXiv:2005.0842}}]. [Erratum: JCAP 05, E01 (2021)].

\bibitem{Shu:2020cjw}
F.-W. Shu, {\it {Vacua in novel 4D Einstein-Gauss-Bonnet Gravity: pathology and
  instability?}},  {\em Phys. Lett. B} {\bf 811} (2020) 135907,
  [\href{http://xxx.lanl.gov/abs/2004.0933}{{\tt arXiv:2004.0933}}].

\bibitem{Mahapatra:2020rds}
S.~Mahapatra, {\it {A note on the total action of 4D Gauss\textendash{}Bonnet
  theory}},  {\em Eur. Phys. J. C} {\bf 80} (2020), no.~10 992,
  [\href{http://xxx.lanl.gov/abs/2004.0921}{{\tt arXiv:2004.0921}}].

\bibitem{Banerjee:2020dad}
A.~Banerjee, T.~Tangphati, D.~Samart, and P.~Channuie, {\it {Quark Stars in 4D
  Einstein\textendash{}Gauss\textendash{}Bonnet Gravity with an Interacting
  Quark Equation of State}},  {\em Astrophys. J.} {\bf 906} (2021), no.~2 114,
  [\href{http://xxx.lanl.gov/abs/2007.0412}{{\tt arXiv:2007.0412}}].

\bibitem{Ge:2020tid}
X.-H. Ge and S.-J. Sin, {\it {Causality of black holes in 4-dimensional
  Einstein\textendash{}Gauss\textendash{}Bonnet\textendash{}Maxwell theory}},
  {\em Eur. Phys. J. C} {\bf 80} (2020), no.~8 695,
  [\href{http://xxx.lanl.gov/abs/2004.1219}{{\tt arXiv:2004.1219}}].

\bibitem{Yang:2020jno}
K.~Yang, B.-M. Gu, S.-W. Wei, and Y.-X. Liu, {\it {Born\textendash{}Infeld
  black holes in 4D Einstein\textendash{}Gauss\textendash{}Bonnet gravity}},
  {\em Eur. Phys. J. C} {\bf 80} (2020), no.~7 662,
  [\href{http://xxx.lanl.gov/abs/2004.1446}{{\tt arXiv:2004.1446}}].

\bibitem{Lin:2020kqe}
Z.-C. Lin, K.~Yang, S.-W. Wei, Y.-Q. Wang, and Y.-X. Liu, {\it {Equivalence of
  solutions between the four-dimensional novel and regularized EGB theories in
  a cylindrically symmetric spacetime}},  {\em Eur. Phys. J. C} {\bf 80}
  (2020), no.~11 1033, [\href{http://xxx.lanl.gov/abs/2006.0791}{{\tt
  arXiv:2006.0791}}].

\bibitem{Yang:2020czk}
S.-J. Yang, J.-J. Wan, J.~Chen, J.~Yang, and Y.-Q. Wang, {\it {Weak cosmic
  censorship conjecture for the novel $4D$ charged Einstein-Gauss-Bonnet black
  hole with test scalar field and particle}},  {\em Eur. Phys. J. C} {\bf 80}
  (2020), no.~10 937, [\href{http://xxx.lanl.gov/abs/2004.0793}{{\tt
  arXiv:2004.0793}}].

\bibitem{Bzowski:2018fql}
A.~Bzowski, P.~McFadden, and K.~Skenderis, {\it {Renormalised CFT 3-point
  functions of scalars, currents and stress tensors}},  {\em JHEP} {\bf 11}
  (2018) 159, [\href{http://xxx.lanl.gov/abs/1805.1210}{{\tt
  arXiv:1805.1210}}].

\bibitem{Coriano:2018bbe}
C.~Corian\`o and M.~M. Maglio, {\it {Exact Correlators from Conformal Ward
  Identities in Momentum Space and the Perturbative $TJJ$ Vertex}},  {\em Nucl.
  Phys.} {\bf B938} (2019) 440--522,
  [\href{http://xxx.lanl.gov/abs/1802.0767}{{\tt arXiv:1802.0767}}].

\bibitem{Bzowski:2013sza}
A.~Bzowski, P.~McFadden, and K.~Skenderis, {\it {Implications of conformal
  invariance in momentum space}},  {\em JHEP} {\bf 03} (2014) 111,
  [\href{http://xxx.lanl.gov/abs/1304.7760}{{\tt arXiv:1304.7760}}].

\bibitem{Bzowski:2015pba}
A.~Bzowski, P.~McFadden, and K.~Skenderis, {\it {Scalar 3-point functions in
  CFT: renormalisation, beta functions and anomalies}},  {\em JHEP} {\bf 03}
  (2016) 066, [\href{http://xxx.lanl.gov/abs/1510.0844}{{\tt
  arXiv:1510.0844}}].

\bibitem{Riegert:1984kt}
R.~J. Riegert, {\it {A Nonlocal Action for the Trace Anomaly}},  {\em Phys.
  Lett.} {\bf 134B} (1984) 56--60.

\bibitem{Gundlach:1999cu}
C.~Gundlach, {\it {Critical phenomena in gravitational collapse}},  {\em Living
  Rev. Rel.} {\bf 2} (1999) 4,
  [\href{http://xxx.lanl.gov/abs/gr-qc/0001046}{{\tt gr-qc/0001046}}].

\bibitem{Coriano:2021nvn}
C.~Corian\`o, M.~M. Maglio, and D.~Theofilopoulos, {\it {The Conformal Anomaly
  Action to Fourth Order (4T) in $d=4$ in Momentum Space}},
  \href{http://xxx.lanl.gov/abs/2103.1395}{{\tt arXiv:2103.1395}}.

\bibitem{Coriano:2017mux}
C.~Corian\`o, M.~M. Maglio, and E.~Mottola, {\it {TTT in CFT: Trace Identities
  and the Conformal Anomaly Effective Action}},  {\em Nucl. Phys.} {\bf B942}
  (2019) 303--328, [\href{http://xxx.lanl.gov/abs/1703.0886}{{\tt
  arXiv:1703.0886}}].

\bibitem{Coriano:2018zdo}
C.~Corian\`o and M.~M. Maglio, {\it {Renormalization, Conformal Ward Identities
  and the Origin of a Conformal Anomaly Pole}},  {\em Phys. Lett.} {\bf B781}
  (2018) 283--289, [\href{http://xxx.lanl.gov/abs/1802.0150}{{\tt
  arXiv:1802.0150}}].

\end{thebibliography}
\providecommand{\href}[2]{#2}\begingroup\raggedright\endgroup

\end{document}